\documentclass[reprint,prl,aps,superscriptaddress]{revtex4-2}
\AtBeginDocument{\RenewCommandCopy\qty\SI}
\usepackage{times}
\usepackage{graphicx}
\usepackage{amsmath, braket, amsfonts}
\usepackage{amssymb}
\usepackage{sidecap}
\usepackage{bm, color, ulem}
\usepackage{xcolor}
\usepackage{ragged2e}
\usepackage{wrapfig,lipsum,booktabs}
\usepackage{lineno}
\usepackage{placeins}

\usepackage{siunitx}
\newcommand{\abs}[1]{\left\lvert #1 \right\rvert}

\newcommand{\be}{\begin{equation}}
\newcommand{\ee}{\end{equation}}
\newcommand{\WSE}{WSe$_2$}

\usepackage{mathtools}
\usepackage{physics}
\newcommand{\normord}[1]{:\mathrel{\mkern2mu #1 \mkern2mu}:}
\newcommand{\hBN}{\textit{h}-BN~}

\DeclareUnicodeCharacter{03BD}{{$\nu$}}

\newcommand{\caltechPH}{Department of Physics, California Institute of Technology, Pasadena, California 91125, USA}
\newcommand{\caltechTH}{Walter Burke Institute for Theoretical Physics, California Institute of Technology, Pasadena, California 91125, USA}
\newcommand{\caltechIQIM}{Institute for Quantum Information and Matter, California Institute of Technology, Pasadena, California 91125, USA}
\newcommand{\UCSB}{Department of Physics, University of California at Santa Barbara, Santa Barbara, California 93106, USA}
\newcommand{\nimsTT}{
Research Center for Materials Nanoarchitectonics, National Institute for Materials Science,  1-1 Namiki, Tsukuba 305-0044, Japan
}
\newcommand{\nimsKW}{
Research Center for Electronic and Optical Materials, National Institute for Materials Science, 1-1 Namiki, Tsukuba 305-0044, Japan}
\newcommand{\harvardPH}{Department of Physics, Harvard University, Cambridge, Massachusetts 02138, USA}

\makeatletter
\def\maketitle{
\@author@finish
\title@column\titleblock@produce
\suppressfloats[t]}
\makeatother

\begin{document}
\title{Superconductivity and spin canting in spin-orbit proximitized rhombohedral trilayer graphene}
\author{Caitlin L. Patterson}
\email{These authors contributed equally} 
\author{Owen I. Sheekey}
\email{These authors contributed equally} 
\author{Trevor B. Arp}
\email{These authors contributed equally} 
\author{Ludwig F. W. Holleis}
\email{These authors contributed equally} 
\affiliation{\UCSB}
\author{Jin Ming Koh}
\affiliation{\caltechPH}
\affiliation{\harvardPH}
\author{Youngjoon Choi}
\affiliation{\UCSB}
\author{Tian Xie}
\affiliation{\UCSB}
\author{Siyuan Xu}
\affiliation{\UCSB}
\author{Evgeny Redekop}
\affiliation{\UCSB}
\author{Grigory Babikyan}
\affiliation{\UCSB}
\author{Haoxin Zhou}
\affiliation{\UCSB}
\email{Current address: Department of Electrical Engineering and Computer Sciences, University of California, Berkeley, California 94720, USA.} 
\author{Xiang Cheng}
\affiliation{\UCSB}
\author{Takashi Taniguchi}
\affiliation{\nimsTT}
\author{Kenji Watanabe}
\affiliation{\nimsKW}
\author{Chenhao Jin}
\affiliation{\UCSB}
\author{\'Etienne Lantagne-Hurtubise}
\affiliation{\caltechPH}
\affiliation{\caltechIQIM}
\author{Jason Alicea}
\affiliation{\caltechPH}
\affiliation{\caltechIQIM}
\affiliation{\caltechTH}
\author{Andrea F. Young}
\email{andrea@physics.ucsb.edu}
\affiliation{\UCSB}
\date{\today}
\maketitle 

\textbf{
Graphene and transition metal dichalcogenide flat-band systems show similar phase diagrams, replete with magnetic~\cite{cao_correlated_2018,
sharpe_emergent_2019,chen_evidence_2019,chen_tunable_2020,li_quantum_2021} and superconducting~\cite{
cao_unconventional_2018, park_tunable_2021,
zhou_superconductivity_2021, zhou_isospin_2022, xia_unconventional_2024, guo_superconductivity_2024} phases. 
An abiding question has been whether magnetic ordering competes with superconductivity or facilitates pairing.  
The advent of crystalline graphene superconductors~\cite{zhou_superconductivity_2021,zhou_isospin_2022,zhang_enhanced_2023,holleis_nematicity_2024,li_tunable_2024} enables a new generation of controlled experiments to probe the microscopic origin of superconductivity.  
For example, recent studies of Bernal bilayer graphene show a dramatic increase in the observed domain and critical temperature $T_c$ of superconducting states in the presence of enhanced spin-orbit coupling~\cite{zhang_enhanced_2023,holleis_nematicity_2024,li_tunable_2024}; the mechanism for this enhancement, however, remains unclear.  
Here, we show that introducing spin-orbit coupling in rhombohedral trilayer graphene (RTG) via substrate proximity effect generates new superconducting pockets for both electron and hole doping, with maximal $T_c\approx$ 300~mK three times larger than in RTG encapsulated by hexagonal boron nitride alone. 
Using local magnetometry and thermodynamic compressibility measurements, we show that superconductivity straddles an apparently continuous transition between a spin-canted state with a finite in-plane magnetic moment and a state with complete spin-valley locking.
This transition is reproduced in our Hartree-Fock calculations, where it is driven by the competition between spin-orbit coupling and the carrier-density-tuned Hund's interaction.
Our experiment suggests that the enhancement of superconductivity by spin-orbit coupling is driven not by a change in the ground state symmetry or degeneracy but rather by a quantitative change in the canting angle.  
These results align with a recently proposed mechanism for the enhancement of superconductivity in spin-orbit coupled rhombohedral multilayers~\cite{dong_superconductivity_2024}, in which fluctuations in the spin-canting order contribute to the pairing interaction.} 

\section{Introduction}
Recent experiments on two-dimensional flat-band systems spanning twisted graphene multilayers, twisted transition metal dichalcogenide (TMD) hetero- and homo-bilayers, and rhombohedral graphene multilayers have revealed persistent similarities in their correlated phase diagrams.  
Most strikingly, all of these systems show an intricate interplay between superconductivity and isospin magnetism, with superconductivity often appearing within or immediately adjacent to phases with reduced isospin degeneracy in the carrier density- and displacement-field-tuned low temperature phase diagram. 
Different multilayer systems share a number of key similarities, including Van Hove singularities within a carrier density range accessible by the electric field effect, as well as valley degeneracy and trigonally warped single-particle Fermi surfaces.    
These underlying similarities account for the observation of phenomena ranging from spin and orbital magnetism and associated quantized anomalous Hall states~\cite{sharpe_emergent_2019, serlin_intrinsic_2020,
shi_electronic_2020, chen_tunable_2020, li_quantum_2021,
anderson_programming_2023, park_observation_2023, foutty_mapping_2024,lu_fractional_2024}, intervalley coherence~\cite{nuckolls_quantum_2023,kim_imaging_2023,arp_intervalley_2024}, 
and other signatures of strong correlations across a variety of microscopically dissimilar systems. 
Also ubiquitous is superconductivity~\cite{
cao_unconventional_2018, park_tunable_2021,
zhou_superconductivity_2021, zhou_isospin_2022, xia_unconventional_2024, guo_superconductivity_2024}; however, no  single mechanism currently accounts for all observations.      
Of course, different experimental systems are distinguished by significant quantitative differences in their single-particle band structure and the relative strength of electron-electron interactions  (including lattice-scale interactions that distinguish spin and valley flavors). Combined with the lack of quantitative predictive power characterizing most theoretical treatments of superconductivity, 
the question remains open as to whether or not superconductivity arises from a common origin in these different platforms.

\begin{figure*}
    \centering
    \includegraphics[width=\textwidth]{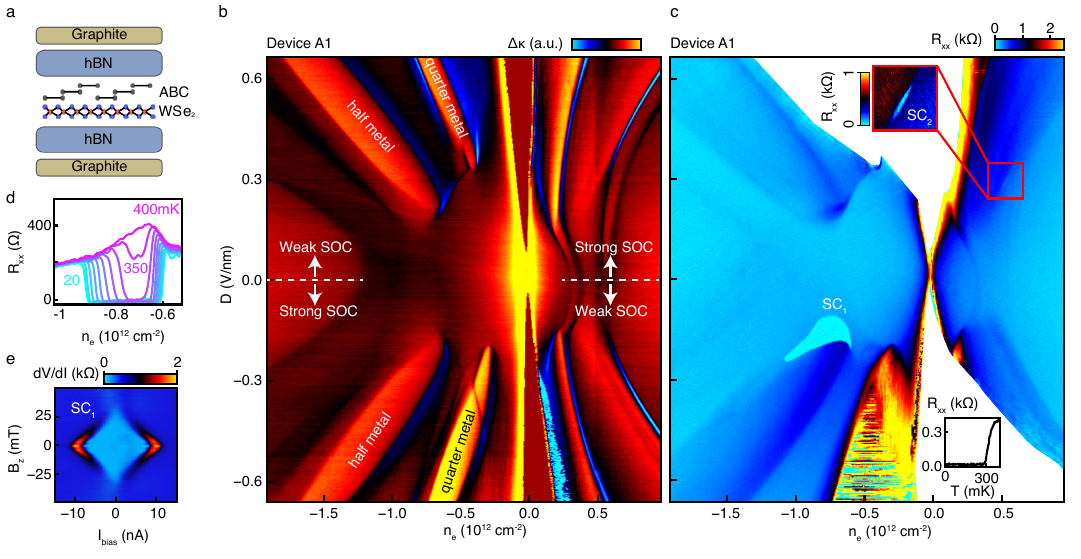} 
    \caption{\textbf{Superconductivity in WSe$_2$-supported rhombohedral trilayer graphene.}
    \textbf{(a)}  
    Schematic of a RTG device encapsulated between hBN and WSe$_2$ flakes. 
    \textbf{(b)} $n_e$- and $D$-dependent inverse compressibility ($\kappa=\partial \mu/\partial n$) at $B=0$~T and $T=20$~mK for Device A1. Both the overall scale and $\kappa=0$ point are left uncalibrated throughout the manuscript, so that data represents the change in inverse compressibility, $\Delta \kappa$.  
    \textbf{(c)} $n_e$- and $D$-dependent longitudinal resistivity ($R_{xx}$) at our base temperature of $T\approx 20$ mK and $B=0$~T.    Regions where high contact resistance precludes accurate resistivity measurements are masked in white.
    Two superconducting pockets are observed, denoted SC$_1$ and SC$_2$, one each for hole- and electron doping.    
    The top inset shows the region containing SC$_2$ with adjusted color scale (see also Extended Data Fig. \ref{fig:electronSC}). 
    The bottom inset shows $R_{xx}(T)$ at $n_e$ = -0.7$\cdot$10$^{12}$cm$^{-2}$ and $D$ = -0.15 V/nm, corresponding to maximal observed superconducting $T_c$.
    \textbf{(d)} Density dependent $R_{xx}$ for several temperatures at $D$ = -0.15 V/nm.
    \textbf{(e)} Current ($I_{\rm bias}$) and magnetic field ($B_z$) dependence of SC$_1$ at the maximum $T_c$ point.} 
    \label{fig:1}
\end{figure*}

Crystalline graphene multilayers offer a reproducible platform for quantitative studies addressing this issue. 
For example, in Bernal bilayer graphene, it was shown that enhancing spin-orbit coupling via proximity to a WSe$_2$ substrate~\cite{wang_origin_2016,island_spinorbit-driven_2019} increases both the domain (in magnetic field, carrier density $n_e$, and applied perpendicular displacement field $D$) and maximum critical temperature of superconductivity~\cite{zhang_enhanced_2023,holleis_nematicity_2024, li_tunable_2024}.  
This result is not intuitive: while spin-orbit coupling preserves the Kramers degeneracy key to superconductivity, it may either raise or lower density of states.  As a result, a systematic increase in $T_c$ with spin-orbit coupling strength is not expected on general theoretical grounds. 
Moreover, under large perpendicular applied electric fields, the intrinsic Kane-Mele spin-orbit coupling already present in graphene systems~\cite{sichau_resonance_2019,banszerus_observation_2020,arp_intervalley_2024} has the same form as the Ising spin-orbit coupling induced by a WSe$_2$~substrate. 
Proximity enhancement of spin-orbit coupling is, consequently, not expected to change the symmetry of the Hamiltonian; rather, a quantitative difference affecting the energetic competition between broken-symmetry states is likely responsible for the appearance of new superconducting pockets. 
Theoretical work to date has presented competing scenarios for the mechanism by which superconductivity is enhanced~\cite{chou_enhanced_2022,jimeno-pozo_superconductivity_2022,wagner_superconductivity_2023,dong_signatures_2023,dong_transformer_2023,curtis_stabilizing_2023,pantaleon_superconductivity_2023,shavit_inducing_2023,dong_superconductivity_2024,son_switching_2024}.

\section{Superconductivity in RTG/WS\MakeLowercase{e}$_2$} 

To explore the origin of spin-orbit enhancement of superconductivity, we study several RTG devices fabricated on WSe$_2$~substrates with geometry shown schematically in Fig.~\ref{fig:1}a (see Fig.~\ref{fig:devices} and the methods for a description of all measured devices). 
As compared to Bernal bilayer graphene, RTG has the advantage that its magnetic phase diagram, at least for hexagonal boron nitride encapsulated samples, is comparatively well understood~\cite{zhou_half-_2021,zhou_superconductivity_2021,arp_intervalley_2024}. 
In particular, past experiments have shown that `bare' RTG---i.e., RTG encapsulated between hBN substrates without an additional WSe$_2$~cladding layer---features quarter- and half-metal phases in which only one or two of the four combined spin- and valley isospin flavors are occupied~\cite{zhou_half-_2021}. 
Figure~\ref{fig:1}b shows the inverse compressibility measured in a WSe$_2$-supported RTG device (Device A1, with Ising spin-orbit coupling strength $\lambda=1.5$~meV, see Fig. \ref{fig:CPABC9_10_Ising}) as a function of $n_e$ and $D$. Throughout the manuscript, we present inverse compressibility in arbitrary units and up to an uncalibrated additive constant, and so denote it $\Delta \kappa$ (calibrated inverse compressibility data on similar systems have been presented elsewhere~\cite{zhou_half-_2021,arp_intervalley_2024}).
The phase diagram closely resembles that of bare RTG~\cite{zhou_half-_2021}.  We label regions of elevated $\Delta \kappa$ associated with half- and quarter-metal phases whose Fermi seas have simply connected topology (s) within each isospin flavor. These benchmark states appear at similar $n_e$ and $|D|$ in both bare and \WSE -supported samples. 
In addition, the compressibility for positive and negative values of $D$ is qualitatively similar, despite the fact that the proximity-induced spin-orbit coupling is expected to be strong only for $n_e\cdot D>0$ \cite{khoo_-demand_2017,gmitra_proximity_2017}, when the carriers are polarized adjacent to the \WSE. 
This observation is consistent with the Fermi surface degeneracy being primarily determined by the interplay of Coulomb interactions and the kinetic energy. 
In this picture, the position of the half- and quarter-metal phases is set by energetic competition at scales comparable to the Fermi energy, $E_F\approx 10$~meV, and is not strongly affected by changes in the spin-orbit coupling strength on the $\lambda\approx 1$~meV scale. 

Transport measurements, however, reveal a striking asymmetry between the strong- and weak $\lambda$ quadrants of the $n_e-D$ plane (see Fig.~\ref{fig:1}c). 
While no superconductivity is observed when carriers are polarized into the weak spin-orbit quadrants ($n_e\cdot D<0$), we find two superconducting pockets---one each for hole and electron dopings, denoted SC$_1$ and SC$_2$, respectively---when carriers are polarized onto the layer adjacent to the WSe$_2$~substrate.  
For SC$_1$, we find maximum $T_c\approx$ 300~mK (see Fig.~\ref{fig:1}d, as well as data from a second, identically fabricated Device A2 in Fig.~\ref{fig:secondSC_device}), while for SC$_2$ we find $T_c\approx$ 40~mK (see Fig.~\ref{fig:electronSC}c).
Both superconductors show current and magnetic field responses (see Figs.~\ref{fig:1}e and \ref{fig:electronSC}d) in line with past experiments on crystalline graphene superconductors~\cite{zhou_superconductivity_2021,zhou_isospin_2022,zhang_enhanced_2023,holleis_nematicity_2024,li_tunable_2024}. 
Specifically, in-plane critical fields (see Fig.~\ref{fig:PVR}) fall both below and above the Pauli limit, exceeding 10 for specific values of $n_e$ and $D$. While critical fields exceeding the Pauli limit are natural with strong Ising spin-orbit coupling, critical fields falling below the Pauli limit may reflect depairing mechanisms from in-plane orbital moments given imperfect layer polarization \cite{holleis_nematicity_2024}.

In bare RTG two superconducting pockets are observed for hole doping~\cite{zhou_superconductivity_2021} and one for electron doping (Fig.~\ref{fig:electronSC}e-h); however, they appear in different regions of the $n_e-D$ plane than in Fig.~\ref{fig:1}. Notably, all of the superconducting pockets in bare RTG are suppressed in Device A for \textit{both} strong and weak spin-orbit coupling.  The regions of the phase diagram showing superconductivity differ qualitatively between bare and spin-orbit-enhanced RTG.  For example, in bare RTG, the most robust superconductor emerges within a phase consistent with single-particle band structure~\cite{zhou_superconductivity_2021}; here, in contrast, SC$_1$ is positioned near the low-$|D|$ end of a half metal phase. 
In addition, the maximal $T_c$ in SC$_1$ is approximately three times higher than for bare RTG~\cite{zhou_superconductivity_2021}.  
These results are reminiscent of Bernal bilayer graphene, where  proximity induced spin-orbit coupling nucleates new, higher-$T_c$ superconducting pockets as compared to bare Bernal bilayers~\cite{zhang_enhanced_2023,holleis_nematicity_2024,li_tunable_2024}.  

\section{Magnetometry and Compressibility}

\begin{figure*}
\centering
\includegraphics[width=183mm]{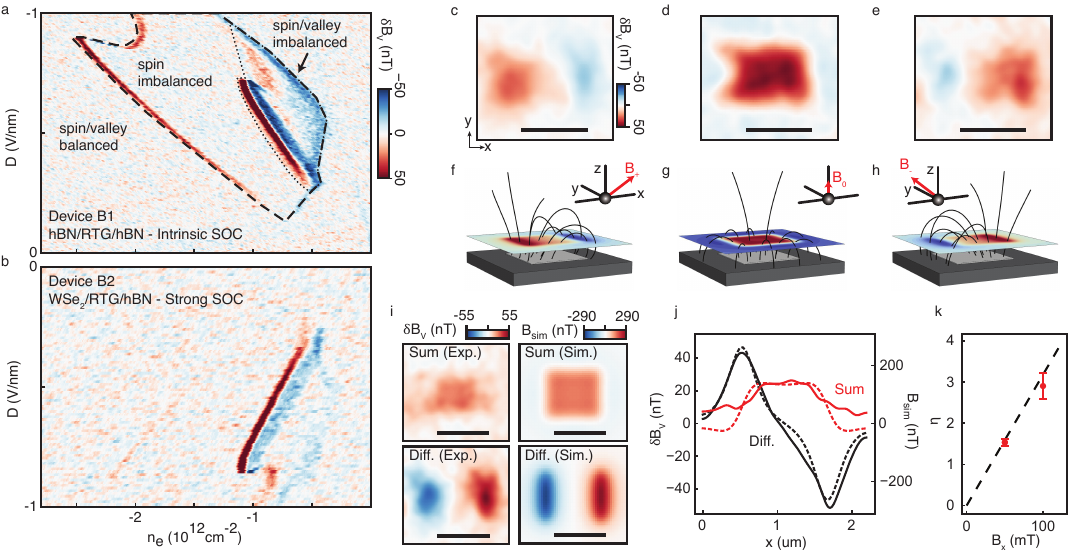}
    \caption{\textbf{Magnetic imaging of symmetry broken states in rhombohedral trilayer graphene.}  
    \textbf{(a)} Magnetic signal $\delta B_V$ in response to a modulation of the bottom gate voltage (see Methods), plotted as a function of $n_{\textrm{e}}$ and $D$ at a single point above Device B1.  
    \textbf{(b)} Same measurement as in panel a over Device B2.  
    \textbf{(c)} Spatial image of $\delta B_V$ over an isolated region of Device B1 for ($n_{\textrm{e}}$, $D$) = (-1.91, -0.61). The scale bar corresponds to 1~$\mu$m. Here $\vec B =\vec B_+\equiv$ (100mT, 0, 46mT).
    \textbf{(d)} The same as in panel c for $\vec B =\vec B_0=$ (0, 0, 46mT).
    \textbf{(e)} The same as in panel c for $\vec B =\vec B_0=$ (-100mT, 0, 46mT).
    \textbf{(f)} Schematic illustration of the flux lines and out-of-plane fringe field above a two dimensional sample with a uniform density of magnetic moments aligned to the applied magnetic field $\vec B=\vec B_+$. 
    \textbf{(g)} The same as in panel f, for $\vec B=\vec B_0$.
    \textbf{(h)} The same as in panel f, for $\vec B=\vec B_-$. 
    \textbf{(i)} Left column: sum and difference of data in panels c and e. 
    Right column: simulated sum and difference assuming a dipole density of $n_e=1.91\times 10^{12}$cm$^{-2}$ 
    (see methods and Extended Data Fig.~\ref{fig:SQUID_Simulation_extended} for additional information). 
    \textbf{(j)} Comparison of measured ($\delta B_v$, solid, left axis) and simulated (dotted, right axis) sum and difference signals across the center of the data shown in panel i. 
    \textbf{(k)} Ratio between the total range of the sum and difference measurements (see main text) for $B_x=$~50~mT and $B_x=$~100~mT.  The dotted line shows the result of simulations assuming magnetic moments align with the applied magnetic field.} 
    \label{fig:2} 
\end{figure*}

One hypothesis for the enhancement of superconductivity with increasing $\lambda$ is that strong spin orbit changes the spin polarization of the symmetry broken phases. 
In bare RTG, the spin structure of the magnetic phases is dominated by the ferromagnetic Hund's coupling. 
This term favors parallel alignment between the physical spins for carriers in opposite valleys, and leads to spin polarization for valley-balanced phases, which include the half-metal as well as intervalley coherent quarter-metals.  
Spin-orbit coupling, by contrast, breaks the SU(2) spin symmetry preserved by Hund's coupling, and favors out-of-plane, \textit{antiparallel} spin alignment for carriers in opposite valleys.

We probe the spin polarization using nanoSQUID-on-tip magnetometry~\cite{vasyukov_scanning_2013}, which allows us to image the local fringe magnetic field arising from the magnetically ordered phases. 
We use an indium SQUID~\cite{anahory_squid--tip_2020} with diameter 141~nm and magnetic field sensitivity of 4~nT/$\sqrt{\textrm{Hz}}$ at $T=$ 300mK; the SQUID is flux-biased to its sensitive point with an applied 46mT magnetic field perpendicular to the sample. 
Figure~\ref{fig:2}a shows measurements of the differential magnetic fringe field at a fixed spatial position $\approx$160~nm above a bare RTG sample.  
The data are acquired by applying a low frequency modulation of $\delta v_b=$ 12~mV to the bottom gate voltage (corresponding to a simultaneous modulation of $\delta n_e= $ 0.018 $\times 10^{12}$~cm$^{-2}$ and $\delta D=$ 0.0016~V/nm). $\delta B_V$ corresponds to the response of a nanoSQUID-on-tip sensor at the same frequency. 
In this contrast mode, signals are visible when the magnetization of the RTG undergoes $n_e$- or $D$-dependent changes larger than our noise floor. 
Comparing $\delta B_{\mathrm{V}}$ to $\Delta \kappa$~\cite{zhou_half-_2021,zhou_superconductivity_2021,arp_intervalley_2024}, we identify three regions of the $n_e$- and $D$-tuned parameter space distinguished by their net magnetization.  
At the highest and lowest densities, the system is spin and valley balanced and no signal is observed.  
A second region, demarcated by a dotted line, shows large $\delta B_{\mathrm{V}}$ with both negative and positive sign, consistent with a valley-imbalanced phase~\cite{zhou_half-_2021,arp_intervalley_2024} where $\delta B_{\mathrm{V}}$ is generated by changes in orbital magnetization~\cite{das_unconventional_2023}. 
A strong $\delta B_{\mathrm{V}}$ signal is also observed at higher density, localized to a narrow strip we identify as the boundary of a spin imbalanced phase (marked by a dashed line), consistent with prior literature~\cite{zhou_superconductivity_2021}. Figure~\ref{fig:2}b shows the fringe magnetic field above  Sample B2, where the RTG is supported by WSe$_2$. 
While the orbital magnetization signals at low density are qualitatively similar, the magnetic signal associated with the spin transition is strongly suppressed, implying a significant change in the nature of the spin imbalanced phase (see Fig.~\ref{fig:squid_full} for additional data). 

We focus first on the spin anisotropy in bare RTG.  
Figures~\ref{fig:2}c-e show the result of three spatially resolved measurements of a micron-sized area of bare RTG (Device B1).  
We measure $\delta B_V$ at fixed $D=-0.61$~V/nm,  $n_e\approx -1.91 \times 10^{12}$cm$^{-2}$ and $B_z=$~46~mT.  The three images show measurements for $B_x= +100$~mT, 0~mT, and $-100$~mT, respectively; we denote these magnetic fields $B_{+}$, $B_0$, and $B_{-}$.
For $B_{\pm}$, the measured fringe fields show opposite sign change across the device, while no such sign change is seen for $B_x=0$.  
These observations are in qualitative agreement with the simulated fringe fields generated by an isotropic spin moment that aligns with the applied magnetic field, shown in Figs. \ref{fig:2}f-h.  

We quantify the degree of spin anisotropy in bare RTG by analyzing the sum and difference of the data measured with applied $B_+$ and $B_-$.  
Figure~\ref{fig:2}i compares these experimentally determined quantities to simulations of the fringe magnetic field of a rectangular sample of dimensions $1.15\mu $m$\times 1.0 \mu$m that is uniformly magnetized in the applied field direction. 
The spatial structure of the measured sum and difference signals agree well with simulations (see also Fig. \ref{fig:2}j), although the signal magnitudes are lower than expected if the spin polarization were complete (see Extended Data Fig.~\ref{fig:SQUID_missing_moments}).
To analyze the data quantitatively, we consider the ratio $\eta$ of the total range of the sum and total range of the difference signals (see Methods).   
$\eta$ effectively measures the magnitude of the in-plane moment relative to the out-of-plane moment. 
Figure~\ref{fig:2}k compares $\eta$ between data and simulations (assuming a fully isotropic spin moment that aligns with the applied magnetic field) for two values of $B_x$.
Based on the agreement between measurements and theory, we conclude that for the $n_e$ and $D$ values used, the spin moment in bare RTG shows no signatures of anisotropy induced by spin-orbit coupling within the limits of our experimental error.

\begin{figure*}
    \centering
    \includegraphics[]{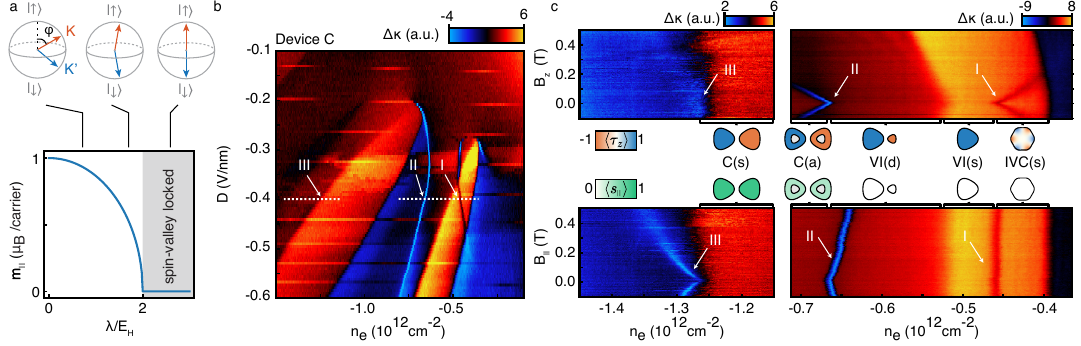}
\caption{
    \textbf{Competition between Hund's and spin-orbit coupling in \WSE-supported RTG.} 
    (\textbf{a}) In-plane magnetic moment $m_\parallel$ of valley-balanced phases as a function of the ratio of Ising spin-orbit coupling strength $\lambda$ and the Hund's energy $E_H$, Eq.~\eqref{eq:canting_evolution}. The Bloch sphere schematics show the configuration of the spin magnetic moment of electrons in the $K$ (red) and $K'$ (blue) valleys.  The canting angle $\varphi$ is defined as shown. 
    For $\lambda\geq 2E_H$,  $\varphi = 0$ and the system is spin-valley-locked (SVL). For $\lambda< 2E_H$, spin rotation symmetry about the $z$ axis is spontaneously broken, and a finite in-plane moment develops. 
    (\textbf{b}) $\Delta \kappa$ as a function of $n_e$ and $D$ for Device C. 
    (\textbf{c}) $\Delta \kappa$ as a function of $n_{e}$ and $B_{z}$ (top row) and $B_{\parallel}$ (bottom row) for $D =$~-0.4~V/nm along the trajectories shown in panel a. 
    The schematics show the Fermi surface topology determined from quantum oscillations, along with their valley polarization, $\langle \tau_{z} \rangle$ and in-plane spin polarization, $\langle s_{\parallel} \rangle$. 
    Phases are labeled by their isospin order and topology as described in the main text. The density ranges indicated are at $\vec B = 0$.}
    \label{fig:3}
\end{figure*}

Naively, the absence of anisotropy in the bare RTG measurements of Fig.~\ref{fig:2} is
surprising, as $\lambda\approx$ 50~$\mu$eV is much 
larger than the Zeeman energy $E_z=g\mu_B |\vec B_\pm | \approx 12.7$~$\mu$eV in our experiment. This absence can be explained by considering the interplay of the spin-orbit and Hund's couplings.  
Spin-orbit coupling breaks the spin SU(2) symmetry; spin polarized  phases must then develop either easy-axis or easy-plane anisotropy. 
Easy-plane phases maintain valley balance but develop a non-zero canting angle between spins in opposite valleys, for an energy gain of order $\lambda^2/E_H$ where $E_H=J_H\cdot n_p$ is the energy of the Hund's coupling. 
Here $n_p$ is the polarization density 
and we estimate the coupling constant as $J_H\approx 2$~meV$ \times 10^{-12}$~cm$^2$~\cite{alicea_graphene_2006, zhou_half-_2021, koh_correlated_2023, arp_intervalley_2024, wei_landau-level_2024} (see Supplementary information). 
Within a Ginzburg-Landau description of spin-canted phases~\cite{dong_superconductivity_2024}, the in-plane magnetization evolves as
\begin{equation}
     m_\parallel \propto \sin (\varphi) = \sqrt{1 - \left( \frac{\lambda}{2 E_H} \right)^2},
     \label{eq:canting_evolution}
 \end{equation} 
where $\varphi$ is the canting angle (see Fig.~\ref{fig:3}a). 
In bare RTG for $n_e \approx n_p \approx 2 \times 10^{12}$~cm$^{-2}$, we have $\lambda/E_H \approx$ 0.01 and $\varphi\approx$ 89.5$^\circ$. The correspondingly small energy gain from spin-orbit coupling is consistent with the low degree of spin anisotropy observed in Fig.~\ref{fig:2}. 

It is tempting to ascribe the suppression of magnetic signal at the spin transition for \WSE-supported RTG in Fig.~\ref{fig:2}b to the disappearance of a net $m_\parallel$ for $\varphi=0$, corresponding to a spin-valley locked state. However, for $\lambda = 1.5$ meV, Eq.~\eqref{eq:canting_evolution} yields $\varphi\approx 79^\circ$ using the same parameters as above, suggesting that appreciable spin canting survives at large polarization densities, even in the presence of \WSE~proximity-induced spin-orbit coupling.  
Indeed, a small residual magnetic signal is visible at the spin transition in Fig.~\ref{fig:2}b, possibly arising from the predicted residual $m_\parallel$. 
Unfortunately, the finite $B_z = 46$~mT required to flux-bias our SQUID allows for the possibility that this signal may arise from the orbital magnetic susceptibility of one or both adjoining phases. 
As the fringe-field signals are too small to allow for detailed spatial characterization, we instead return to measurements of the inverse compressibility to search for evidence of spin canting.

Figure~\ref{fig:3}b shows a phase diagram of $\Delta \kappa$ versus $n_{e}$ and $D$ for a third WSe$_2$-supported RTG device (Device C). As for Device A, $|\lambda|\approx 1.5$~meV  for Device C, although with opposite sign (see Fig. \ref{fig:CPABC9_10_Ising}). 
We focus on three phase boundaries, labeled I, II, and III in Fig.~\ref{fig:3}b, characterized by negative compressibility features we associate with first-order phase transitions.  
The response of a first-order phase transition line to in- and out-of-plane magnetic fields is given by the Clausius Clapeyron-type relation $dn^*/dB=\Delta m/\Delta \mu$, where $n^*$ is the density at the phase transition, $\Delta m$ is the jump in magnetization in the direction of the applied field $B$, and $\Delta \mu$ is the chemical potential jump. 
Figure~\ref{fig:3}c shows the $B_z$ and $B_\parallel$ dependence of $\Delta \kappa$ along the trajectories shown in Fig.~\ref{fig:3}b.  
Boundary I depends strongly on $B_z$, allowing us to identify it with a boundary separating a valley imbalanced state (VI) at higher hole density and an intervalley coherent (IVC) state at lower density (both with simple (s) Fermi surface topology)---identical to observations in RTG~\cite{arp_intervalley_2024}. However, in contrast to bare RTG where the IVC state is spin polarized, boundary I is insensitive to $B_\parallel$.  This insensitivity implies that the IVC state has no net in-plane moment, as expected for a spin-valley-locked IVC phase~\cite{koh_correlated_2023, zhumagulov_emergent_2023},  
i.e., with canting angle $\varphi=0$ (see also Extended Data Fig.~\ref{fig:SVLIVCtheory} and supplementary Fig.~\ref{fig:SVLIVCexperiment}).
Boundary III corresponds to the spin transition discussed in Fig.~\ref{fig:2} for bare RTG that separates the canted spin phase at low $|n_e|$ from a spin unpolarized phase at higher $|n_e|$~\cite{zhou_superconductivity_2021}.  
A cusp-like behavior is observed as a function of $B_\parallel$, suggesting that the low $|n_e|$ phase is spin polarized and thus that spin canting indeed survives in \WSE -proximitized RTG.  
The low $|n_e|$ boundary of the spin canted state is identified with boundary II. Here, the higher $|n_e|$ state is favored by $B_\parallel$ but disfavored by $B_z$ relative to the valley-imbalanced state at lower $|n_e|$, whose orbital moment couples linearly to $B_z$. 
The schematic depictions of the Fermi surfaces in Fig.~\ref{fig:3}c summarize our understanding of the competing phases in this regime.  
For each identified phase, we classify its Fermi surface topology as either annular (a), simple (s) or disjoint (d); in addition, we show the expected in-plane spin polarization $s_\parallel$ and out-of-plane valley polarization $\tau_z$ as a function of wave vector $\vec k$ within the occupied (hole) states.

\begin{figure}
\centering
    \includegraphics[width=\columnwidth]{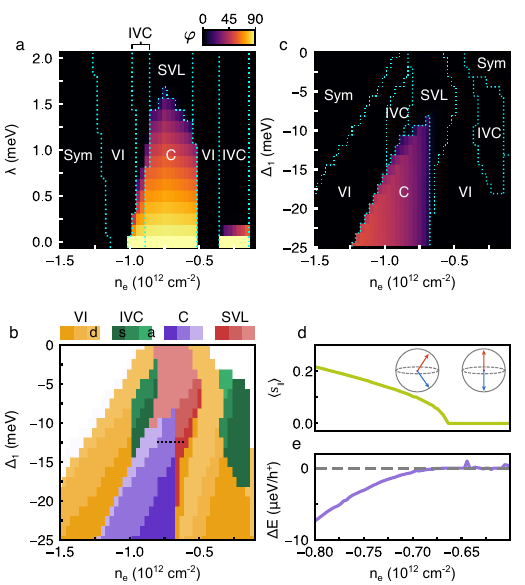}
    \caption{ \textbf{Self-consistent Hartree-Fock calculations.}
    (\textbf{a}) Ground state and canting angle at fixed interlayer potential $\Delta_{1} = -10$~meV as a function of $n_e$ and Ising spin-orbit coupling strength $\lambda$.  
    Boundaries between phases with distinct isospin symmetry are marked, and the color scale represents the canting angle $\varphi$ (see Methods and supplementary Fig.~\ref{fig:HF_Lambda_Theory} for details).
    (\textbf{b}) Calculated ground state versus $n_e$ and $\Delta_{1}$ with $\lambda = 1.5$ meV. The isospin structure (VI, IVC, C, and SVL) is coded by color, while 
    the Fermi surface topology is encoded by shading with `a' for annular, `s' for simply connected, and `d' for disjoint. 
    White regions denote spin-valley-locked ground states
    that resemble 
    the single-particle band structure prediction, which we denote `Sym'.  
    (\textbf{c}) Canting angle $\varphi$ of the ground state orders shown in panel \textbf{b}. 
    (\textbf{d}) Expectation value of the in-plane spin moment, $\langle s_\parallel \rangle$, along the trajectory shown in panel b at $\Delta_{1} = -12.5$~meV. The order parameter evolves continuously from zero, as expected at a continuous phase transition.  
    \textbf{e} The energy difference per carrier, $\Delta E$, between C and SVL phases, again showing a continuous evolution.}
    \label{fig:4}
\end{figure}

\section{Hartree-Fock simulations}

Evidently, changing the density at fixed $\lambda$ and $D$ allows us to tune across the phase boundary where spin canting develops. 
This identification is supported by self-consistent Hartree-Fock calculations (see Ref.~\onlinecite{koh_correlated_2023} and the Supplementary Information for details of implementation).  
Figure~\ref{fig:4}a shows the evolution of the Hartree-Fock phase diagram as a function of $n_e$ and $\lambda$ for interlayer potential difference $\Delta_1 = -10$~meV. 
As in the experiment, the simulations show a density-tuned alternation between valley-imbalanced phases and valley balanced phases within which we plot the canting angle $\varphi$. At $\lambda = 0$ the quarter-metal IVC phase and half-metal valley-balanced phase are both spin polarized ($\varphi = 90^{\circ}$).  Both phases develop a canting angle $\varphi <  90^{\circ}$ with increasing $\lambda$; however, just as observed in the experiment, the transition to a spin-valley-locked phase with $\varphi=0$ occurs at much lower $\lambda$ for the low density IVC order. At $\lambda = 1.5$ meV, the simulations capture the intermediate canting angle of the half metal phase as well as $\varphi= 0$ observed for the IVC phase (see Extended Data Fig.~\ref{fig:SVLIVCtheory} for IVC details).  

Figure~\ref{fig:4}b shows the calculated $n_e$-$D$ phase diagram for $\lambda=1.5$ meV.  
The simulation reproduces several features of the experimental phase diagram, including the alternating pattern of disjoint, simple, and annular Fermi surface topology as well as the density-tuned transition from VI to IVC order within the quarter metal state. 
Notably, it also shows a transition between a spin-canted phase at high $|\Delta_1|$ and a spin-valley-locked phase at low $|\Delta_1|$. Figure~\ref{fig:4}c shows $\varphi$ across the same parameter regime as in Fig.~\ref{fig:4}b; $\varphi$ increases with increasing $|n_e|$ in the half-metal regime, consistent with the free-energy prediction of  Eq.~\eqref{eq:canting_evolution}.
Within Hartree-Fock, the onset of spin canting out of the spin-valley-locked phase 
appears continuous, as illustrated in Fig.~\ref{fig:4}d where we plot the in-plane spin expectation value $\langle s_\parallel\rangle$ evaluated in the ground state.
At the transition $\langle s_\parallel\rangle$ smoothly tends to zero, as does the energy difference between the spin-canted and (symmetric) spin-valley-locked phases (Fig.~\ref{fig:4}e). This behavior is consistent with similar numerical results for spin-orbit proximitized Bernal bilayer graphene~\cite{koh_symmetry-broken_2024}.

\section{Spin canting and superconductivity}
\begin{figure*}
    \centering
    \includegraphics[width=\textwidth]{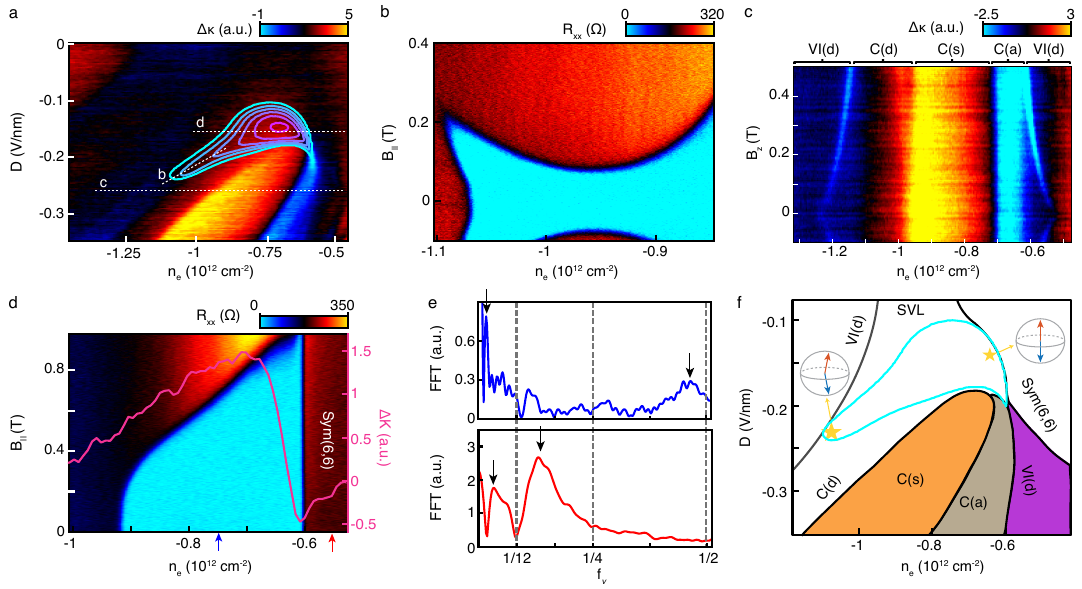}
    \caption{\textbf{Interplay of spin canting and superconductivity in Device A.} 
    \textbf{(a)} $n_e-D$ phase diagram of $\Delta \kappa$ in the vicinity of SC$_1$ in Device A1. 
    The contours represent constant superconducting $T_c$ = 20mK, 100mK, 150mK, 200mK, 250mK, and 285mK.
    The dashed lines indicate along which direction the data of panels b and d are taken. 
    \textbf{(b)} In-plane magnetic field $B_\parallel$ dependence of SC$_1$ along along the corresponding line in panel a. 
    In-plane field induced superconductivity is observed at the high density end.
    \textbf{(c)} $\Delta \kappa$ along the trajectory marked in panel a.   
    \textbf{(d)} in-plane magnetic field $B_\parallel$ dependence cutting across SC$_1$ at $D$ = -0.162 V/nm. 
    Overlaid in pink is a measurement of $\Delta \kappa$ at $B_\parallel$ = 0.0T.
    \textbf{(e)} Fourier transform FFT of quantum oscillations measurements (see Fig. \ref{fig:QO}) at the points marked by the blue and red arrow in panel d.
    The black arrows mark the frequencies corresponding to the large and small Fermi pockets.    
    \textbf{(f)} Schematic phase diagram around SC$_1$ summarizing the experimental results.
    The Bloch sphere schematics illustrate spin canting from $\varphi =$~0 to finite canting with 0 $< \varphi <$~90$^{\circ}$ within the domain of SC$_1$.}
    \label{fig:5}
\end{figure*}

Our thermodynamic measurements establish the survival of spin-canted order in \WSE -supported RTG, at least at large $D$, in agreement with Hartree-Fock simulations.  
These simulations further suggest that spin canted order gives way to spin valley locking via a continuous transition at lower $D$.  However, we have not established the experimental relation between the domain where superconductivity is observed and the spin canting transition.  
To this end, Fig.~\ref{fig:5}a shows $\Delta \kappa$ overlaid with contours of constant $T_c$ for the SC$_1$ pocket (see additional resistivity data in Fig. \ref{fig:QO}a), along with three trajectories in the $n_e$-$D$ plane.
Figure~\ref{fig:5}b shows the dependence of $R_{xx}$ on in-plane magnetic field along a  trajectory  (shown in panel a) that crosses the high-$|D|$ end of the superconducting pocket.  Notably, the domain of superconductivity \textit{expands} with increasing $B_\parallel$.  
This highly unusual phenomenology is most easily explained if the superconducting state has a finite $m_\parallel$ relative to the adjoining state.  
$\Delta \kappa$ measurements indeed show a signature of 
a first-order transition (see Fig.~\ref{fig:SC_spincanting}) at higher $B_\parallel$.  
To constrain the nature of the competing phase, Fig.~\ref{fig:5}c shows $B_z$-dependent measurement of $\Delta \kappa$ along the trajectory marked in panel a.
 Apparently, the competing phase at higher $|n_e|$ has finite spontaneous out-of-plane moment consistent with a valley imbalanced phase, as evidenced by the cusp in the first-order phase transitions relative to the canted spin phases.  
These data suggest that the superconducting state, near its highest $|D|$ extension, emerges from a phase with finite $m_\parallel$, consistent with spin-canted order. 

At lower $|D|$,  in contrast, the superconducting domain does not expand with applied $B_\parallel$ (Fig.~\ref{fig:5}d).  Simultaneous measurements of $\Delta \kappa$ show that the low $|n_e|$ boundary of the superconducting region coincides with a first-order phase transition, marked by a negative dip in $\Delta \kappa$. 
Quantum oscillation measurements (see Figs.~\ref{fig:5}e  and Fig.~\ref{fig:QO}) show that the superconducting state occurs in a regime of reduced degeneracy with two large and two small Fermi pockets, while on the low $|n_e|$ side of the transition the Fermi surfaces are consistent with the single-particle band structure expectation of two sets of six identically sized pockets split by the Ising spin-orbit coupling.
A superconductor in a spin-canted normal state is expected to expand relative to this spin-valley locked competing phase (which we denote Sym(6,6)); we thus conclude that the canting angle $\varphi$ at this boundary is experimentally indistinguishable from zero.   

Figure~\ref{fig:5}f summarizes the key experimental findings gleaned from the data presented in Fig.~\ref{fig:5}a-e as well as from additional quantum oscillation and tilted field measurements presented in Extended Data Fig.~\ref{fig:QO} and \ref{fig:SC_spincanting}.  Superconductivity occurs entirely within a region of disjoint Fermi sea topology, with both minority and majority Fermi surfaces.  
At the low $|D|$ end, proximity-induced Ising spin-orbit coupling dominates and the ground state has $\varphi\approx$~0.  Within the superconducting region, a canted phase with 0~$< \varphi <$~90$^{\circ}$ develops.  Notably, no sharp features are observed in $\Delta \kappa$ within the superconducting region (see Fig.\ref{fig:5}a and \ref{fig:QO}b), consistent with the continuous transition predicted by our Hartree-Fock simulations.  

\section{Discussion}

Our results empirically show that the nucleation of new superconducting pockets by spin-orbit proximity effect is likely generic to rhombohedral graphene multilayers.
Moreover, proximity of the superconducting phase to the spin-canting transition motivates a detailed comparison of the phenomenology to the theoretical prediction that superconductivity may arise via spin-fluctuation-induced pairing~\cite{dong_superconductivity_2024}. 
Most obviously, superconductivity occurs directly atop an apparently continuous transition where canting order develops.  Superconductivity further arises only in regimes hosting both large and small Fermi pockets, with no superconductivity observed where the Fermi surfaces have simple topology. 
This result also accords with the predictions of Ref.~\cite{dong_superconductivity_2024}, in which scattering between large and small pockets plays a central role.  Perhaps the most dramatic result of our experiment, however, is that canting order itself is present both with and without proximity induced spin-orbit coupling in the same regime of $D$ and $n_e$, with similar Fermi surfaces (see Extended Data Fig.~\ref{fig:bare_rtg_cd}), but superconductivity arises at observable temperatures only when $\lambda$ is large.  This finding again accords with the prediction that the pairing interaction is maximized for $\varphi\ll \pi/2$.  

It is interesting to compare spin canting fluctuations to fluctuations of an IVC order parameter, which have been proposed to mediate superconductivity in both twisted and crystalline graphene systems~\cite{kozii_spin-triplet_2022, chatterjee_inter-valley_2022, dong_signatures_2023}. 
Indeed, the recent observation of intervalley coherence as the parent state of superconductivity in twisted bilayer~\cite{nuckolls_quantum_2023} and trilayer~\cite{kim_imaging_2023} graphene, in combination with the present results, may point towards a universal pairing mechanism at play in graphene superconductors, relying on the low-energy collective modes mandated by the spontaneous breaking of a continuous symmetry.

\textbf{Acknowledgements.} The authors would like to acknowledge discussions with Zhiyu Dong and Erez Berg.
Work in the Young lab at UCSB was primarily supported by the Department of Energy under 
Award DE-SC0020043 to A.F.Y., with additional support provided by the Gordon and Betty Moore Foundation under award GBMF9471.
CLP acknowledges support by the Department of Defense (DoD) through the National Defense Science and Engineering Graduate (NDSEG) Fellowship Program.
T.A., O.S., and  E.R. acknowledge direct support by the National Science Foundation through Enabling Quantum Leap: Convergent Accelerated Discovery Foundries for Quantum Materials Science, Engineering and Information (Q-AMASE-i) award number DMR-1906325; the work also made use of shared equipment sponsored by under this award. 
C.J. acknowledges support from Air Force Office of Scientific Research under award FA9550-23-1-0117. T.X. acknowledges the support from the National Science Foundation Graduate Research Fellowship under Grant No.2139319.
K.W. and T.T. acknowledge support from the JSPS KAKENHI (Grant Numbers 21H05233 and 23H02052) and World Premier International Research Center Initiative (WPI), MEXT, Japan.
J.M.K. is grateful for support from the Agency for Science, Technology and Research (A*STAR) Graduate Academy, Singapore. \'E.L.H. was supported by the Gordon and Betty Moore Foundation’s EPiQS Initiative, Grant GBMF8682. The U.S. Department of Energy, Office of Science, National Quantum Information Science Research Centers, Quantum Science Center supported the high-performance computing component of this work.

\textbf{Author contributions. }
CLP fabricated Devices A and C with the help of YC and XC. 
SX and TX fabricated Device B, supervised by CJ. 
GB, ER fabricated the nanoSQUID on tip sensors. 
TX and HZ fabricated Device D.  
HZ measured Device D.  
TBA and OIS developed the nanoSQUID on tip microscope, acquired the magnetometry measurements on Device B, and analyzed the magnetometry data. 
CLP and YC performed transport and capacitance experiments on Devices A and C. CLP and YC initiated the analysis of the transport and capacitance data from Devices A and C. LFWH performed capacitance measurements of Device B, and completed the data analysis of Devices A and C.  TT and KW provided the hexagonal boron nitride crystals. 
JMK, ELH and JA developed the theoretical model and analysis; self-consistent Hartree-Fock simulations were performed by JMK. 
OIS, TBA, LFWH, ELH, and AFY wrote the paper.  
All coauthors reviewed the manuscript prior to submission.

\bibliographystyle{science}
\bibliography{references}


\clearpage
\newpage
\pagebreak

\onecolumngrid

\begin{center}
\textbf{\large Supplementary information }\\[5pt]
\end{center}

\setcounter{equation}{0}
\setcounter{figure}{0}
\setcounter{table}{0}
\setcounter{page}{1}
\setcounter{section}{0}
\makeatletter
\renewcommand{\theequation}{S\arabic{equation}}
\renewcommand{\thefigure}{S\arabic{figure}}
\renewcommand{\thepage}{\arabic{page}}


\section{Materials and Methods}

\subsection{Sample Fabrication}

In this work we present data from three RTG/WSe$_2$ samples and one sample fully encapsulated by hBN. The latter is the same sample as in previous work~\cite{zhou_superconductivity_2021,zhou_half-_2021,arp_intervalley_2024} with fabrication details outlined there.
For all samples, the exfoliated trilayer graphene flakes were characterized by Raman microscopy~\cite{lui_imaging_2011}.
Rhombohedral regions were then isolated using local anodic oxidation techniques with an atomic force microscope~\cite{huang_electrical_2018}. 
Subsequently, the RTG flakes where assembled for the three different devices as follows:\\

\textbf{Device A, C}: the device is stacked in two steps. 
First, a graphite bottom gate on hBN is picked up with a poly(bisphenol A carbonate) film, and WSe$_2$ is picked up subsequently. The stack is then flipped using a gold coated PDMS stamp following the technique developed in ref.~\cite{kim_imaging_2023}.
This leaves a pristine WSe$_2$/hBN surface serving as the bottom half for the top half of the stack to be placed on.
The top half consisting of top gate, hBN, and RTG is stacked sequentially with a flat PDMS stamp at pickup angles $< 1$ degree between the PDMS stamp and the SiO$_2$ substrate. 
All flakes are picked up between 40 - 110~$^{\circ}$C, with the RTG  picked up at 70~$^{\circ}$C.\\

\begin{figure}
    \centering
    \includegraphics{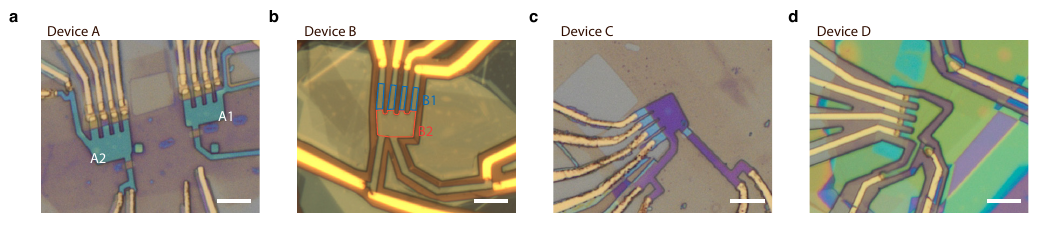}
    \caption{\textbf{Optical micrographs of the different RTG devices:} images of Devices A-D. The scale bar is 10 $\mu m$ in each picture. 
    Data from each device is labeled correspondingly in each figure. 
    All devices details are summarized in table \ref{table:samples}.}
    \label{fig:devices}
\end{figure}

\textbf{Device B}: The bottom half without WSe$_2$ is stacked, but not flipped and instead the poly carbonate stacking film is dissolved in Chloroform and subsequently the sample is heated to 375~$^{\circ}$C in vacuum for several hours to remove polymer residues. 
The top half with WSe$_2$ is then assembled analogous to Device A, C. 
We want to point out that not the full dual gated RTG area is proximitized to WSe$_2$. 
Instead, regions close to the contacts only see hBN substrates on both the top of the bottom which is illustrated in Fig \ref{fig:devices}b.
We denote the different regions as Device B1 (without WSe$_2$) and B2 (with WSe$_2$).
This leads to a superposition of features from both the areas with and without WSe$_2$  support as seen in Fig. \ref{fig:squid_full}c. 
s
Optical micrographs of Devices A-D are shown in Fig. \ref{fig:devices} and sample details are summarized in table \ref{table:samples}.

\begin{center}
\begin{table}[htb]
\begin{tabular}{ |c| c |c| c|c|}
\hline
 \textbf{Sample} & \textbf{SOC $\lambda$} & \textbf{Top/Bottom hBN thickness (nm)}  & \textbf{Main Text Figures} & \textbf{Extended Data Figures} \\
 \hline
 Device A1 & 1.5 meV (see Fig. \ref{fig:CPABC9_10_Ising})& 35 / 35 &   \ref{fig:1},\ref{fig:5} & \ref{fig:CPABC9_10_Ising},\ref{fig:electronSC},\ref{fig:PVR},\ref{fig:SVLIVCtheory}, \ref{fig:QO},\ref{fig:SC_spincanting}, \ref{fig:SVLIVCexperiment} \\
 \hline
 Device A2 & 1.5 meV (see Fig. \ref{fig:CPABC9_10_Ising}) & 35 / 35 &  - & \ref{fig:secondSC_device} \\
 \hline
 Device B1 & 50 $\mu$eV \cite{arp_intervalley_2024} & 8 / 10 &  \ref{fig:2} & \ref{fig:SQUID_Simulation_extended}, \ref{fig:SQUID_missing_moments},\ref{fig:squid_full} \\
 \hline
 Device B2 & undetermined & 8 / 10 &  \ref{fig:2} & \ref{fig:squid_full}, \ref{fig:SVLIVCexperiment} \\
 \hline 
 Device C & $\sim$ -1.5 meV (see Fig. \ref{fig:CPABC9_10_Ising})& 25/10  &  \ref{fig:3} & \ref{fig:SVLIVCexperiment} \\
 \hline
 Device D & 50 $\mu$eV \cite{arp_intervalley_2024} & 25 / 40  &  - & \ref{fig:electronSC},\ref{fig:bare_rtg_cd} \\
 \hline
\end{tabular}
\caption{Device parameters}
\label{table:samples}
\end{table}
\end{center}

\subsection{Electronic transport and compressibility measurements}

Transport measurements were performed using standard lock-in amplifier techniques at a low frequency $< 100$~Hz with the frequency chosen to minimize noise.
Penetration field capacitance measurements of the inverse compressibility were performed using a High Electron Mobility Transistor (HEMT) at base temperature as described in Refs.~\onlinecite{zibrov_emergent_2018, zhou_half-_2021} with a decoupling capacitor and 4K cyrogenic amplifier as described in Ref.~\onlinecite{holleis_nematicity_2024}. 
All transport and capacitance measurements were performed in a dilution refrigerator equipped with a superconducting vector magnet and cryogenic low-pass filters to reduce electron temperatures. 
Unless otherwise specified, all transport and compressibility measurements were performed at base temperature of $\lesssim 20$~mK.

\subsection{NanoSQUID on Tip}

The nanoSQUID on Tip (nSOT) measurements were performed using an indium SQUID fabricated on the end of a pulled quartz pipette using the self-aligned fabrication technique \cite{anahory_squid--tip_2020}. This nSOT had a field period of 132~mT which gives an effective diameter of 141~nm. The nSOT signal was read out by measuring the current through the tip with a series SQUID array amplifier in feedback mode. The sensitivity of the nSOT was measured to be $\approx$ 4~nT/$\sqrt{\textrm{Hz}}$ \cite{anahory_squid--tip_2020}. The nSOT was mechanically controlled using a piezoelectrically excited quartz tuning fork in a phase-locked loop positioned over the sample with attocube nanopositioners to control the tip above the RTG layer. We note that different measurements displayed in this paper were taken under different fixed height conditions, detailed below. Thickness of hBN is determined optically, and `Approximate thickness of stack above RTG' represents the sum of the pickup hBN and the hBN between the top gate and RTG. 

\begin{center}
\begin{table}[htb]
\begin{tabular}{ |c| c |c| c |c|}
\hline
 \textbf{Figures} & \textbf{Height above surface} & \textbf{Stack above RTG} & \textbf{Height above RTG}  \\
 \hline
 Fig.~\ref{fig:2}a, Ext. Fig.~\ref{fig:SQUID_missing_moments} & 110nm & $53 \pm 5$nm & 163nm \\
 \hline
 Fig.~\ref{fig:2}e-k, Fig.~\ref{fig:squid_full}a,b, Fig.~\ref{fig:SQUID_Simulation_extended}a-c  & 150nm & $53 \pm 5$nm  & 203nm \\
 \hline
\end{tabular}
\caption{nSOT scanning parameters}
\label{table:nsot_heights}
\end{table}
\end{center}

As described in the main text the sample was measured by applying a AC excitation to the bottom gate (typically 0.5-3.0~kHz chosen to minimize noise) with an RMS amplitude of 12~mV at the same time as a DC voltage to the top and bottom gates. Spatial scanning data indicates that the ferromagnetic phase transitions, i.e. the contrast in Fig.~\ref{fig:2}a,b, take place in all active regions of the device simultaneously with respect to gate voltage, which is consistent with previous nSOT measurements of RTG \cite{arp_intervalley_2024}. All nSOT measurements were performed in a Helium-3 cryostat with the sample in vacuum, at the base temperature of 300~mK.

\subsection{Field simulations and analysis}
\label{sec:field_analysis}

\begin{figure*}
\centering
\includegraphics{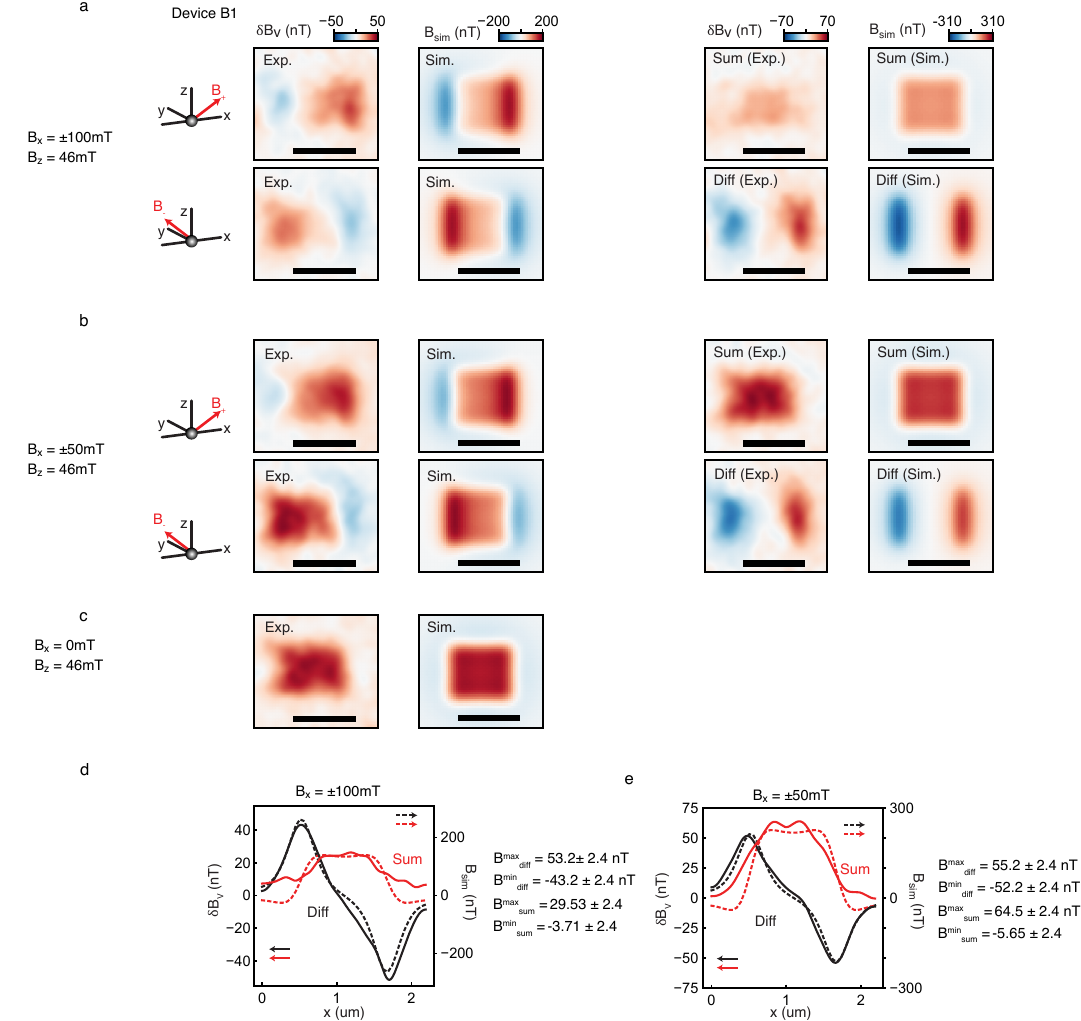}
    \caption{\textbf{
    Comparison of measured and simulated fringe magnetic fields.}
    \textbf{(a)} Experimental and simulated fringe fields for $B_z = 46$mT, $B_x = \pm 100$mT.  Also shown are the sum and difference of experimental and simulated data for opposite applied $B_x$. 
    \textbf{(b)} Same data as in panel a, but for  $B_z = 46$mT and $B_x = \pm 50$mT.
    \textbf{(c)} Experimental and simulated data for $B-x=0$, $B_z=46mT$.  
    \textbf{(d)} Comparison of simulated and experimental data across the centerline of of the scan range for $B_x=\pm 100mT$.  
    \textbf{(e)} 
The same as panel d, for $B_x=\pm 50 mT$. }
    \label{fig:SQUID_Simulation_extended}
\end{figure*}

\begin{figure*}
    \centering
    \includegraphics{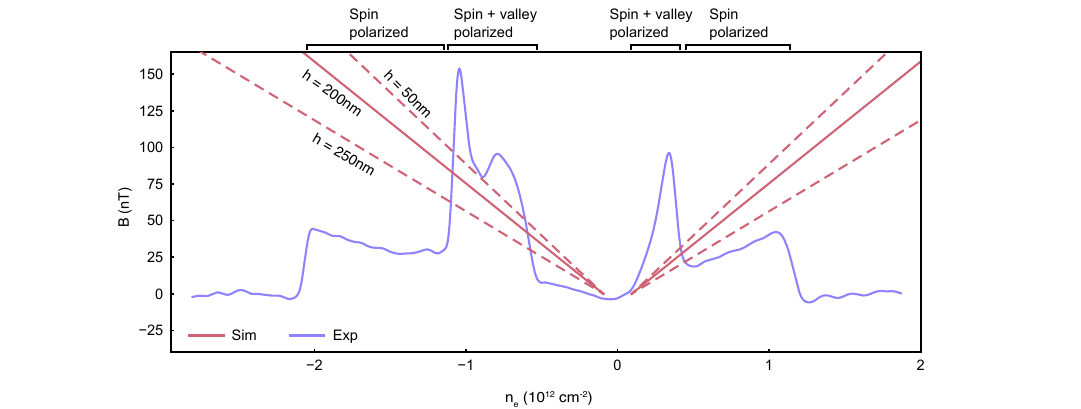}
    \caption{\textbf{Quantitative comparison of magnetic fringe fields above RTG}.
    Measured magnetic fringe field above Device B1 (bare RTG) as a function of density at a fixed $D = -0.677$~V/nm, $B_z=46mT$ and $B_x=B_y=0$. 
    Data are acquired by integration of the measured signal, 
    $B=\int \frac{\delta B_V}{\delta v_b} dv_B$ along a trajectory of constant $D$.  
The data are compared with the simulated fringe field for a two dimensional ferromagnet polarized out of plane with a spin density given by $|n_e|$. 
We include the quantum capacitance effect of a single particle gap of 50~mV for reference. Three different heights above the sample are simulated and notated forming the gray range of possible signal. We note the large discrepancy between the measured signal and the expectation within the spin polarized, valley unpolarized phases, where these numbers are expected to agree if the system is fully spin polarized. 
In experiment, we observe no effect of frequency, or AC excitation amplitude on the magnitude of the measured signal; furthermore, we note that the discrepancy differs for electron and hole doping in regions both expected to be fully polarized.}
    \label{fig:SQUID_missing_moments}
\end{figure*}

To simulate the fringe fields (as shown in the main text (Fig.~\ref{fig:2}) and Extended Data (Fig.~\ref{fig:SQUID_missing_moments}, Fig.~\ref{fig:SQUID_Simulation_extended}) we approximate the sample as a grid of dipoles in the $z = 0$ plane occupying a $1.15 \mu m \times 1.0 \mu m$ rectangle, with the dipole strength adjusted to ensure a given dipole moment density. 
The direction of the dipoles is taken to be uniform and in the direction of the applied magnetic field, and the fringe magnetic field computed in the plane of the SQUID sensor. 

Parameters for all the simulations shown in the text can be found below. 

\begin{center}
\begin{table}[htb]
\begin{tabular}{ |c| c |c| c |c|}
\hline
 \textbf{Figures} & \textbf{Simulated height} & \textbf{Simulated magnetic moment} & \textbf{Effective spin density} & \textbf{Dipole spacing}  \\
 \hline
 Fig.~\ref{fig:SQUID_missing_moments} & 50-350 nm & Variable & Variable & 26 nm\\
 \hline
 Fig.~\ref{fig:2}h-j, Fig.~\ref{fig:SQUID_Simulation_extended}a-c  & 200 nm & 177.134 nA  & $1.91 \times 10^{12}$ cm$^{-2}$& 40 nm\\
 \hline
\end{tabular}
\caption{Magnetic simulation parameters.}
\label{table:simulation_conditions}
\end{table}
\end{center}

While these simulations can be directly compared to our experimentally acquired differential fringe field spatial maps ($\delta B_v$, as in Fig~\ref{fig:2}), the total magnetic moment for the best match to the data does not have a direct physical meaning. This is because $\delta B_v$ does not encode the total fringe field, only the change in fringe field in response to a small voltage modulation to the gate. The spatial imaging presented in this paper is acquired at a gate-tuned transition into the spin-polarized phase in RTG, where our small voltage modulation on the gate leads to a large change in the fringe field. In general, the width and precise location (in gate voltage space) of such a transition will vary with applied in-plane field, as is evident in the behavior of transition III in Fig.~\ref{fig:3}c.  As a result, quantitatively, $\delta B_v$ maps taken at different in-plane field but the same density and displacement field should not be directly compared. 

To allow us to compare datasets at different in-plane fields, we develop a straightforward metric, $\eta$, derived from ratio of the sum and difference images described in the main text. Specifically, we define 
\begin{equation}
    \eta \equiv (B_\text{diff}^\text{max}-B_\text{diff}^\text{min})/(B_\text{sum}^\text{max}-B_\text{sum}^\text{min})\label{eta}
\end{equation}
Fig.~\ref{fig:SQUID_Simulation_extended} shows the values of $B_\text{diff}^\text{max}$, $B_\text{sum}^\text{max}$, $B_\text{diff}^\text{min}$, and $B_\text{sum}^\text{min}$ used for calculation of $\eta$ in the main text.
$\eta$ is plotted in Fig~\ref{fig:2} of the main text for two different in-plane magnetic fields and compare it to simulations uniform moments aligned with the applied magnetic field.

Error bars in $\eta$ are determined by the effective noise in a single `effective pixel' of our nanoSQUID data.  The effective pixel size is set by our spatial resolution, which in turn is set by the SQUID diameter and standoff distance between RTG and SQUID layer.  This is determined by (1) the effective dwell time of on a single pixel and (2) the effective field sensitivity of the nanoSQUID sensor. Raw simulation and experimental data (which is oversampled relative to this size) are Gaussian filtered, with $2\sigma =$~141~nm corresponding to the nanoSQUID diameter. Our magnetic field field noise with finite in-plane field is slightly larger than without, $S_B\approx 10 nT/\sqrt{\text{Hz}}$. The spatial scans presented in the paper represent a roughly 17 by 14 pixel grid, for approximately 221 total pixels with a dwell time per pixel of $\sim$ 17 seconds. This corresponds to standard error on each pixel of $\pm 2.4$~nT. This error is propagated through Eq. \eqref{eta} to generate the error bars shown in Fig. \ref{fig:2}d. 

\subsection{Determination of $\lambda$}

\begin{figure}
    \centering
    \includegraphics{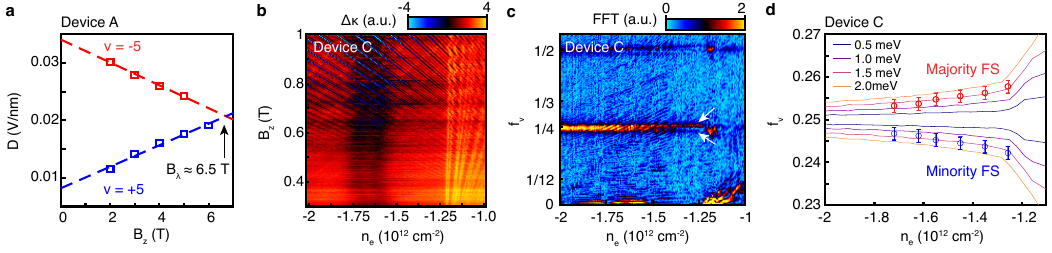}
    \caption{\textbf{Measurement of $\lambda$ in Devices A and C.} 
    \textbf{(a)} Ising extraction from Landau level transitions close to $D \approx 0$~V/nm in Device A1. We plot the displacement field value of the transitions of the $\nu$ = $\pm$5 Landau levels as a function of $B_z$. We linearly extrapolate to where $D_{\pm5}$ are equal and find $B_\lambda$ = 6.5~T which results in $\lambda\approx$ 1.5~meV (for details see methods and equivalent analysis in \cite{island_spinorbit-driven_2019}).
    \textbf{(b)} Quantum oscillations of $
    \Delta\kappa$ at $D$ = 0~V/nm in Device C.
    \textbf{(c)} FFT of the data in panel b revealing a splitting of the peak at $f_\nu=1/4$ due to the Ising SOC, marked by the white arrows.
    \textbf{(d)} comparison of the experimental splitting due to Ising SOC of the 4-fold degenerate Fermi surface (dots) to single particle band structure calculations (lines) for different values of $\lambda$ from 0.5 to 2.0~meV. 
    The error bars represent the full width at half-maximum of the peaks in panel c.
    From this analysis, we find $\lambda \approx$ 1.5~meV in Device C (see Methods).}
    \label{fig:CPABC9_10_Ising}
\end{figure}

Ising Spin-orbit coupling (SOC) leads to a splitting in energy between states where the spins are (anti-) aligned with the orbital degree of freedom.
Hence, there are several experimental signatures which help to determine the Ising SOC $\lambda$.
Landau level (LL) crossings within the lowest LL provide on such tool for determining $\lambda$, and are the preferred method in Bernal bilayer graphene\cite{island_spinorbit-driven_2019}. 
For Device A, we apply this same methodology to our RTG/WSe$_2$ sample.
As the lowest LL in RTG is 6-fold degenerate, the first and last LL to be filled are the $\nu$ = $\pm$ 5 symmetry broken quantum Hall states.  
Analogous to BBG, isospin (or equivalently layer) transitions occur with displacement field $D$.
Because the effective Zeeman splitting arising from the SOC is oriented out of plane, an out-of-plane extrinsic Zeeman splitting will precisely cancel it when $2E_z = 2g\mu_B\cdot B_z = \lambda$, where g is the Land\'e g-factor and $\mu_B$ is the Bohr magneton~\cite{island_spinorbit-driven_2019}.
We show the transition points as a function of both $D$ and $B_z$ in Fig. \ref{fig:CPABC9_10_Ising}a.
Fitting lines to both the $\nu$ = $\pm$ 5 transitions results in $B_\lambda = 6.5$~T, equivalent to $\lambda \approx 1.5$~meV.  

Unfortunately, the above method fails for one sign of $\lambda$\cite{wang_quantum_2019}, expected in half the devices given the random orientation between RTG and \WSE layers.  This is the case in Sample C, so we instead utilize quantum oscillation measurements to determine $\lambda$.  
Quantum oscillations in $\Delta \kappa$ as a function of density and $B_z$ at $D$ = 0 V/nm are shown in Fig. \ref{fig:CPABC9_10_Ising}b.
The normalized frequency $f_\nu$ of the dominant peaks in Fourier transformation gives the size of the Fermi surfaces as a fraction of the total charge volume.
The Ising SOC manifests itself as a splitting of the 4-fold degenerate Fermi surface (FS) states into two slightly larger FS in which spin and valley are aligned and two slightly smaller FS in which they are anti-aligned. 
This splitting is clearly seen in the FFT in Fig. \ref{fig:CPABC9_10_Ising}c (indicated by the white arrows) and grows with decreasing $|n_e|$, as expected. 
We compare this splitting with single particle band structure calculations in Fig.~\ref{fig:CPABC9_10_Ising}d for different values of $\lambda$ from 0.5 to 2.0 meV and find $\lambda \approx$ 1.5 meV to match closest with the experimental data.
The error bars give an estimate of the experimental peak width in the FFT.
We conclude that Device A and C have comparable $\lambda$.

\begin{figure*}
    \centering
    \includegraphics{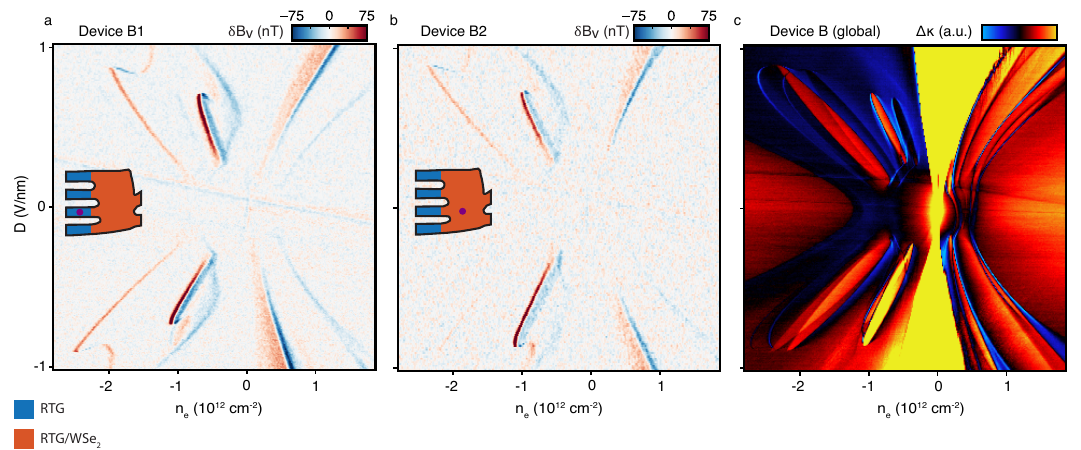}
    \caption{\textbf{Full range RTG and RTG-WSe$_2$~Phase Diagrams of Device B}. 
    \textbf{(a)} Differential magnetism $\delta B$ versus $n_{\textrm{e}}$ and $D$ over the Device B1. The positive displacement field data is the same as Fig.~\ref{fig:1}a. Faint diagonal lines at low $D$ arise from the magnetism of the top graphite gate. 
    \textbf{(b)} $\delta B_V$ in Device B2. The ($n_{\textrm{e}}<0$, $D>0$) and ($n_{\textrm{e}} > 0$, $D < 0$) quadrants are proximitized towards the WSe$_2$~and are shown in Fig.~\ref{fig:1}d.
    \textbf{(c)} Corresponding compressibility phase diagram measured globally, i.e. over both Device B1 and B2. The two regions result in a doubling of many features which are slightly shifted due to the addition of \WSE to the gate dielectric for the B2 region.
    All data in a,b are taken with $B_{z} = 46$~mT and $B_{\parallel} = 0$. Data in c is taken with no field applied.}
    \label{fig:squid_full}
\end{figure*}

Device B has the unique feature that it has regions of hBN+hBN (B1) and WSe$_2$+hBN (B2) encapsulation with large and small SOC. 
However, $\kappa$ -  a global probe of the whole device - is sensitive to both the B1 and B2 areas. 
As the hBN dielectrics used in this device are thin (see table \ref{table:samples}) and WSe$_2$ is approximately 0.7~nm thick, the geometric capacitances between the B1 and B2 regions differ  significantly. 
This leads to an apparent ``doubling'' of many of the half- and quarter-metal features in Fig. \ref{fig:squid_full}c compared to Fig. \ref{fig:1}b.
This precludes us from utilizing either the LL transitions or quantum oscillations to determine $\lambda$ in region B2.
Nevertheless, we want to point out that all three Devices A, B2 and C with Ising SOC look qualitatively similar in the regime of quarter metals (Fig. \ref{fig:SVLIVCexperiment}) where ``doubling'' features do not obscure our measurements. 
Lastly, we want to point out that the intrinsic SOC in RTG devices encapsulated by hBN was recently determined in Device D (see Ref.~\cite{arp_intervalley_2024}), and should be similar in Device B1 as well.

\section{Extended Data}

\begin{figure}
    \centering
    \includegraphics{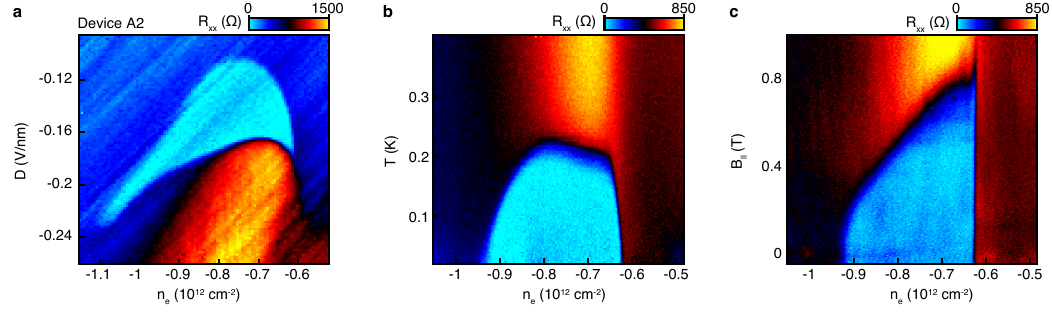}
    \caption{\textbf{Superconductivity in Device A2.} 
    \textbf{(a)} $n_e$- and $D$- dependent $R_{xx}$ near  SC$_1$ in Device A2. 
    \textbf{(b)} Temperature dependence at $D$ = -0.165`V/nm.
    \textbf{(c)} $B_\parallel$ dependence at the same displacement field. 
    All data taken at 1~nA.}
    \label{fig:secondSC_device}
\end{figure}

\begin{figure*}
    \centering
    \includegraphics[width = \textwidth]{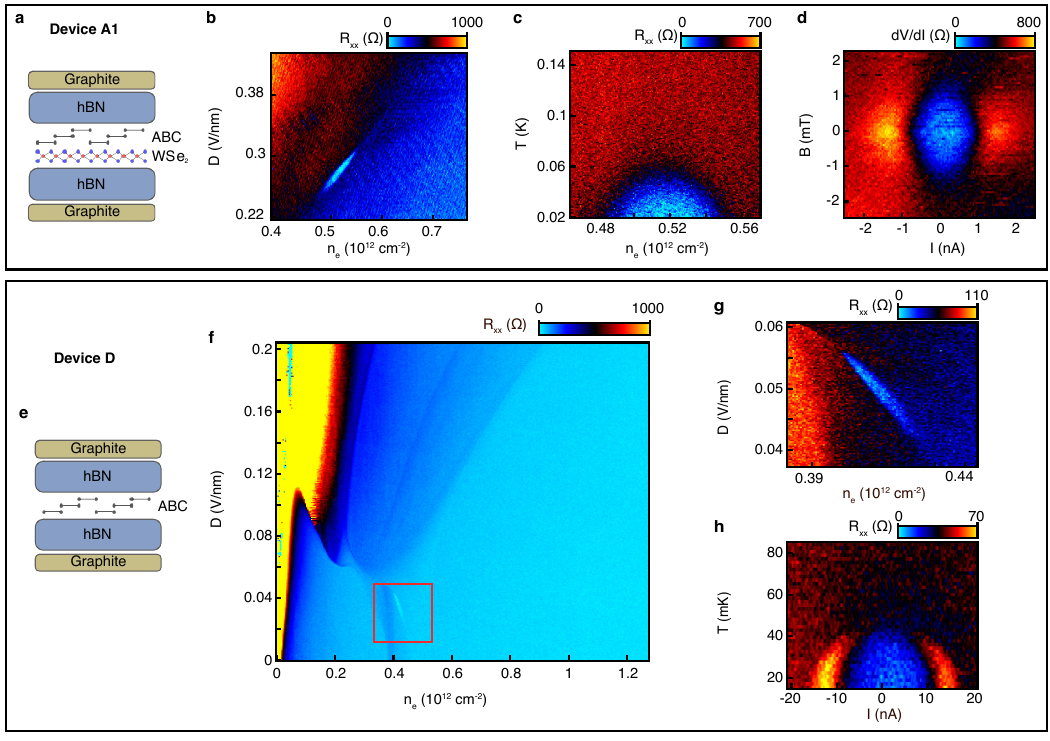}
    \caption{\textbf{Electron side superconductivity with and without \WSE.} 
    \textbf{(a)} Schematic of Device A1. 
    \textbf{(b)} $n_e$-$D$ phase diagram of $R_{xx}$ on the electron-doped side showing a superconductor SC$_2$ Device A1. 
    Same data as Fig.~\ref{fig:1}b, inset. 
    \textbf{(c)} Temperature dependence at $D$ = 0.28 V/nm for Device A1.
    \textbf{(d)} Current ($I$) and $B_z$ dependence at $n_e$ = 0.52$\cdot$10$^{12}$cm$^{-2}$ and $D$ = 0.28 V/nm for Device A1.
    \textbf{(e)} Schematic of an hBN-supported Device D. 
    \textbf{(f)} Electron-doped $n_e$ and $D$ phase diagram of the longitudinal resistance $R_{xx}$ in Device D. 
    \textbf{(g)} Detail near the superconducting region indicated by the red box in panel f. 
    \textbf{(h)} Temperature and current ($I$) dependence at $n_e$ = 0.41$\cdot$10$^{12}$cm$^{-2}$ and $D$ = 0.052 V/nm in Device D. }
    \label{fig:electronSC}
\end{figure*}

\begin{figure*}
    \centering
    \includegraphics[width = \textwidth]{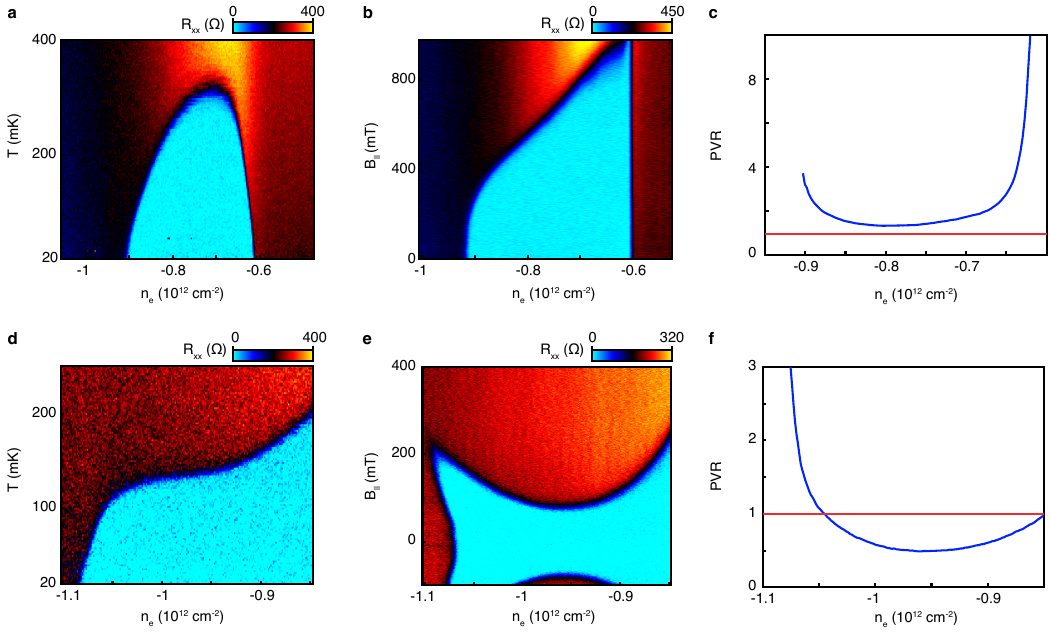}
    \caption{\textbf{Pauli limit violation ratio (PVR) of SC$_1$ in Device A1.} 
    \textbf{(a)} Temperature dependence of $R_{xx}$ at $D$ = -0.162 V/nm, measured with constant alternating current of 1 nA.  
    \textbf{(b)} $B_\parallel$ dependence of $R_{xx}$ at at $D$ = -0.162 V/nm, measured with constant alternating current of 1 nA.
    \textbf{(c)} Pauli limit violation ratio ($PVR$) derived from panels a and b.  The red line represents the Pauli limit, $PVR$ = 1, equivalent to $k_B T_C/(\mu B_C)=1.23$. 
    \textbf{(d)} Temperature dependence of $R_{xx}$ on the $n_e$- and  $D$- dependent trajectory shown in main text Fig.~\ref{fig:5}b. Measured with constant alternating current of 0.5 nA.
    \textbf{(e)} $B_\parallel$ dependence of $R_{xx}$ on the $n_e$- and  $D$- dependent trajectory shown in main text Fig.~\ref{fig:5}b.  Measured with constant alternating current of 1 nA.
     \textbf{(f)} Pauli limit violation ratio ($PVR$) derived from panels d and e.  The red line represents the Pauli limit, $PVR$ = 1, equivalent to $k_B T_C/(\mu B_C)=1.23$. 
    }
    \label{fig:PVR}
\end{figure*}

\begin{figure}
    \centering
    \includegraphics{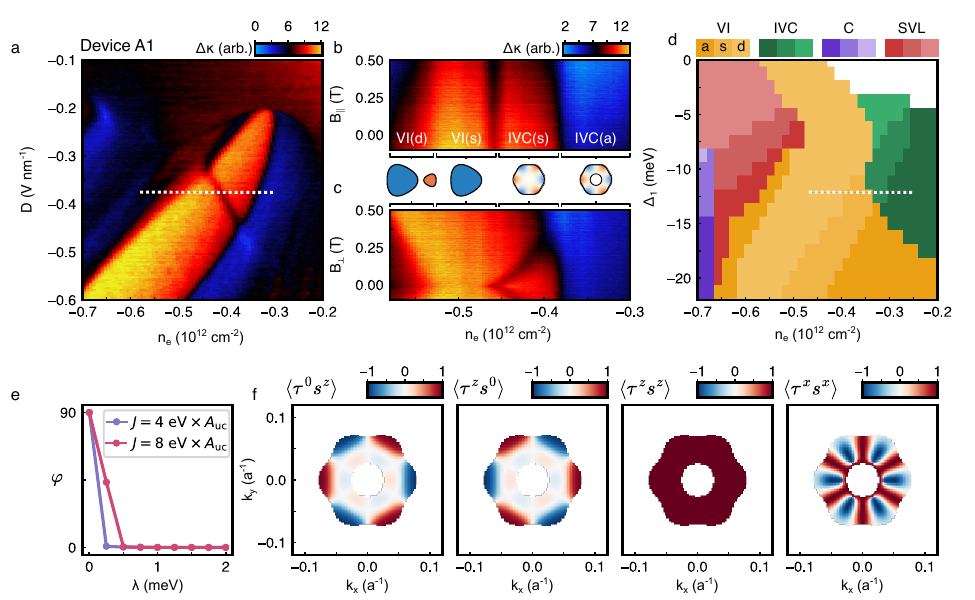}
    \caption{\textbf{Spin/valley locked intervalley coherent quarter metal.} 
    (\textbf{a}) $\Delta \kappa$ as a function of $n_e$ and $D$ for Device A1 in the quarter metal regime. 
    (\textbf{b}) $B_\parallel$ and (\textbf{c}) $B_\perp$ 
    dependence respectively of the phase transitions along the dashed line in panel a. 
    Labels show assignments of Valley Imbalanced (VI) and Intervalley Coherent (IVC) phases based on quantum oscillations, magnetic susceptibility and nSOT magnetometry. Schematics show the fermiology and valley polarization (color coded as in Fig.~\ref{fig:3}c of the main text).  The indicated ranges  correspond to the phase boundaries at $B = 0$. 
    Note that the transition between VI(s) and IVC(s) is $B_{\parallel}$-independent, consistent with spin-valley locking in the IVC phase. Similar data for Devices B2 and C is shown in the supplementary material,  Fig.~\ref{fig:SVLIVCexperiment}).
   (\textbf{d}) Phase diagram determined using self-consistent Hartree-Fock calculations using the parameters from Fig.~\ref{fig:4}b of the main text: $\lambda = $~1.5~meV, $J_H \approx 210$~meV~nm$^2$ ($J_H = 4$eV $\times A_{\rm uc}$), and $\epsilon_r = 30$ (see supplementary information for details). Data show the ground state as a function of $n_e$ and the interlayer potential $\Delta_{1}$, with the isospin order denoted by the color and the Fermi surface topology (a, s or d) indicated by the shading. Here, weakly interacting SVL phases (where the Ising spin splitting is dominated by band structure effects) are rendered in white. The white dashed line is taken at a value of $\Delta_1$ which cuts through the sequence of phases highlighted in panel a.
(\textbf{e}) Evolution of the canting angle $\varphi$ within the quarter-metal IVC regime shown in panel d ($n_e = -0.25 \cdot 10^{12}$~cm$^{-2}$, $\Delta_{1} = -10$~meV) as a function of induced Ising SOC $\lambda$. Canting order disappears in favor of spin-valley locking (i.e. $\varphi = 0$) for much weaker values of $\lambda$ as compared to the half-metal regions. 
    (\textbf{f}) Momentum-resolved textures of several order parameters that characterize the SVL-IVC phase. These data were obtained at $n_e = -0.66 \cdot 10^{12}$~cm$^{-2}$, $\Delta_{1} = -40$~meV, where the Fermi sea has annular topology. 
    }
    \label{fig:SVLIVCtheory}
\end{figure}

\begin{figure*}
    \centering
    \includegraphics[width = \textwidth]{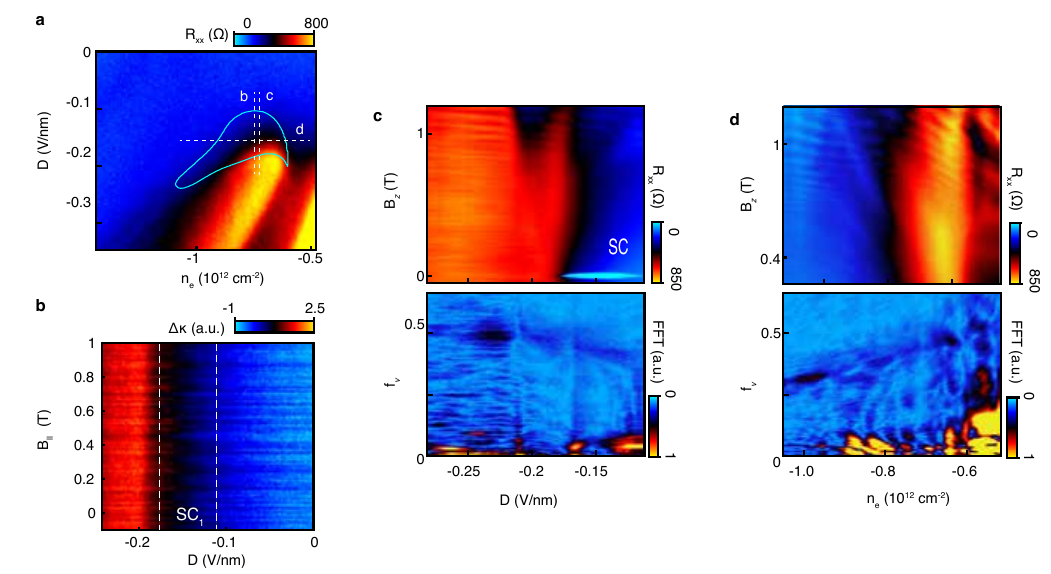}
    \caption{\textbf{Quantum oscillations and Fermiology in the vicinity of SC$_1$ for Device A1.} 
    \textbf{(a)} $\Delta \kappa$ as a function of $n_e$ and $D$ near  SC$_1$.
    \textbf{(b)} $D$- and $B_{\parallel}$ dependence of $\Delta\kappa$ at $n_e$ = 0.81 10$^{12}$cm$^{-2}$  (trajectory b in panel a).  No signature of a  first order phase transition is observed, and no $B-\parallel$ dependence is observed, indicating that superconductivity terminates for this trajectory at a Lifshitz transition without a change in isospin symmetry.
    \textbf{(c)}  Top panel: $R_{xx}$ as a function of $D$ and $B_z$ along the trajectory marked `c' in panel a.  
    Bottom panel: the Fourier transform of $R_{xx}(1/B_z)$.  The frequency $f_\nu$ is normalized by the total charge carrier density.
    \textbf{(d)} The same as in panel c, but for the trajectory marked `d' in panel a.         
       ).  
    }
    \label{fig:QO}
\end{figure*}

\begin{figure*}
    \centering
    \includegraphics[width = \textwidth]{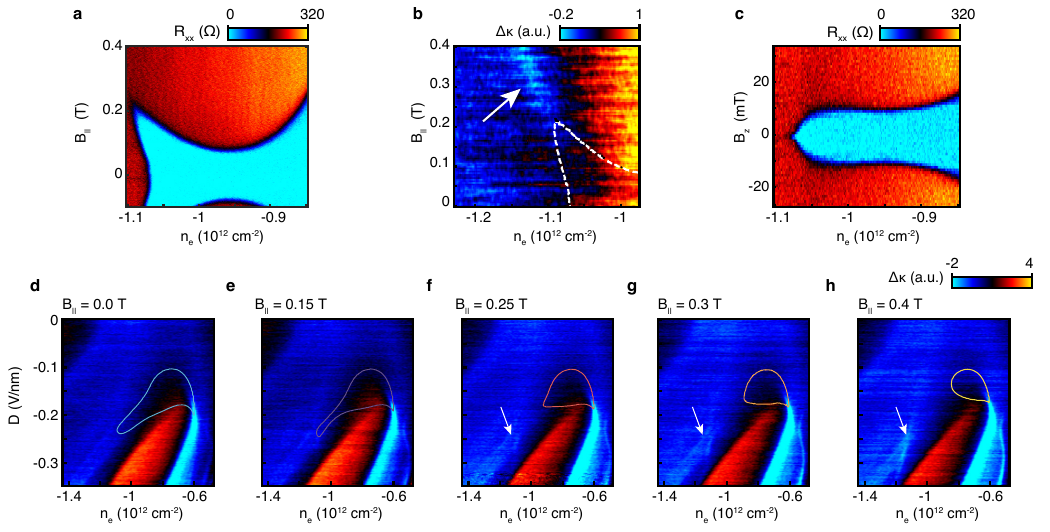}
    \caption{
    \textbf{Spin-canting in the superconducting state.} 
    \textbf{(a)} $B_\parallel$ dependence of $R_{xx}$ in Device A1, identical to panel Fig. \ref{fig:5}b of the main text.  
    \textbf{(b)} $\Delta \kappa$ measured for the same parameters.  A weak dip in $\Delta\kappa$, interpreted as a weakly first order phase transition, appears for $B_\parallel>0.2T$ and is indicated  by the white arrow.   The superconducting region at zero field is demarcated by the dashed contour.
    \textbf{(c)} $B_z$ dependence along the same $n_e$- and $D$-tuned trajectory as in panels a and b. 
    \textbf{(d)-(h)} $n_e$-$D$ maps of $\kappa$ for increasing in-plane field and finite out-of-plane field of 50~mT (d,e), 75~mT (f,g) and 100~mT (h). 
    The superconducting contours measured with 0.5 nA AC current excitation are overlaid. 
    White arrows mark the same the same transition highlighted in panel b.}
    \label{fig:SC_spincanting}
\end{figure*}

\begin{figure*}
    \centering
    \includegraphics[width = \textwidth]{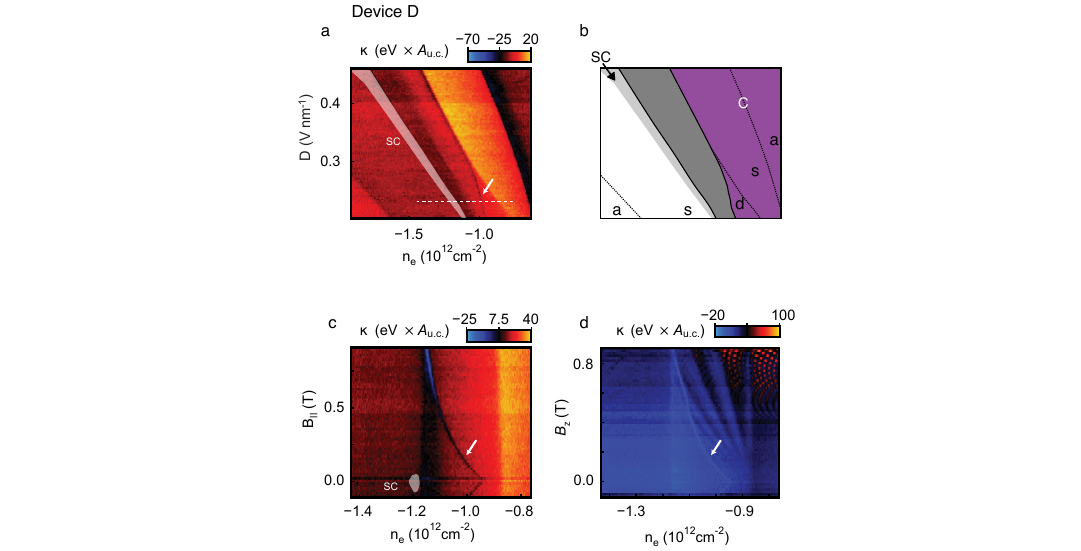}
\caption{\textbf{Evidence for a canted spin phase with disjoint fermi sea topology in bare RTG.} 
\textbf{(a)} $\kappa$ measured in Device D.  The superconducting region is shown in white. 
\textbf{(b)} Proposed phase diagram based on quantum oscillation and relative susceptibility measurements. Purple denotes the canted spin phases (described as `spin polarized' in Ref. \cite{zhou_half-_2021}.  White denotes symmetric phases associated with the single particle band structure.  Gray denotes a spin unpolarized phase with disjoint fermi sea topology. 
\textbf{(c)} $B_\parallel$ dependence along the trajectory shown in panel a.  The cusped transition line confirms the finite in-plane moment of the C(d) phase.  \textbf{(d)} 
Out of plane field dependence over the same trajectory. 
 The similar behavior of the transition line is consistent with an isotropic spin moment, as discussed in the main text. Two oscillation frequencies are visible in the quantum oscillations, confirming the disjoint topology of the fermi seas in this phase.}
    \label{fig:bare_rtg_cd}
\end{figure*}

\clearpage
\section{Supplementary Figures}
\FloatBarrier

\begin{figure}
    \centering
    \includegraphics{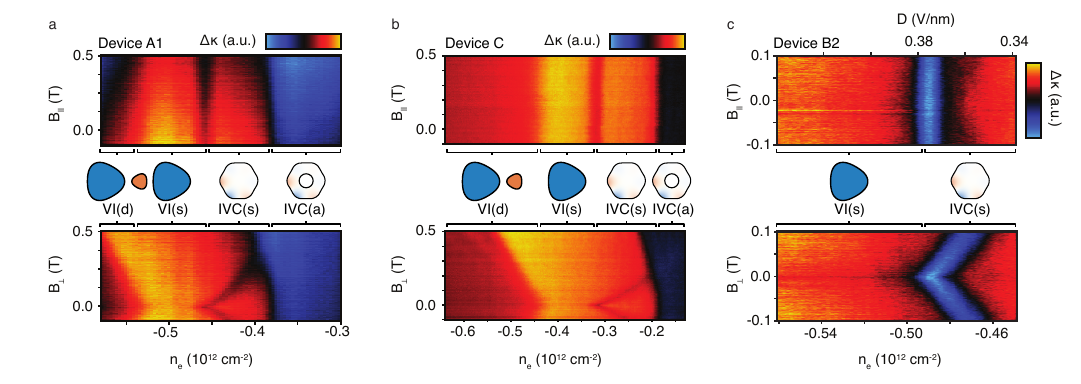}
    \caption{\textbf{Field dependence for all RTG-WSe$_{2}$ devices} 
    In-plane (top row) and out-of-plane (bottom row) field dependence of quarter metal  phase transitions for Devices A1, B2 and C. The curvature of the phase transitions versus field gives information on the relative magnetic moments via a Clausius Clapeyron-type relation, $dn^*/dB=\Delta m/\Delta \mu$, where $n^*$ is the density of the phase transition, $\Delta m$ is the jump in magnetization in the direction of the applied $B$, and $\Delta \mu$ is the chemical potential jump. This when combined with fermiology from quantum oscillations (Ref, \cite{zhou_half-_2021, arp_intervalley_2024}, Supplementary Fig.~\ref{fig:QO}) and nanoSQUID-on-Tip magnetometry (Ref. \cite{arp_intervalley_2024}, Fig.~\ref{fig:squid_full}) identify nature of the Fermi surfaces present shown as the middle row of schematics with color indicating valley polarization $\langle \tau_{z} \rangle$ in the same style as Fig.~\ref{fig:3}c. Phases are identified with labels of their isospin order with topology in parenthesis; the density ranges indicated are at $B = 0$.
    (\textbf{a}) Field dependence for Device A1 along $D =$~-0.374~V~nm$^{-1}$.
    (\textbf{b}) Field dependence for Device C along $D =$~-0.4~V~nm$^{-1}$.
    (\textbf{c}) Field dependence for Device B2 for a trajectory in $n_{e}$ and $D$ across the marked phases.
    The transitions are all broadly consistent across devices, in particular the IVC(s) to IVC(a) and VI(s) to IVC(s) transition show invariance to $B_{\parallel}$ consistent with a spin-valley locked IVC state.
    }
    \label{fig:SVLIVCexperiment}
\end{figure}

\begin{figure}
    \centering
    \includegraphics{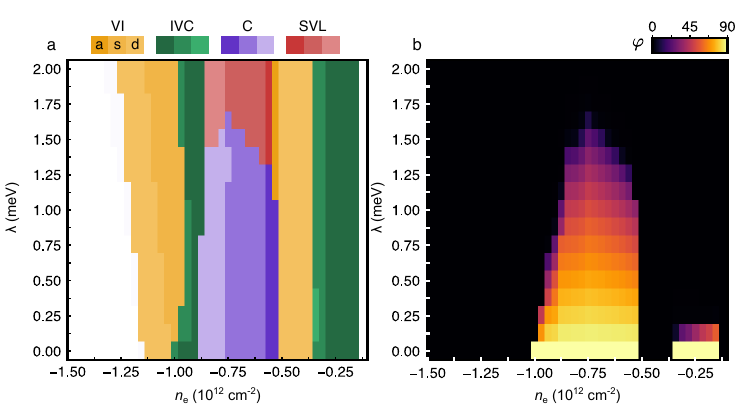}
    \caption{\textbf{Hartree-Fock as a function of $\lambda$.} Self-consistent Hartree-Fock calculations with the same parameters as Fig.~\ref{fig:4}a. 
    (\textbf{a}) Calculated ground state versus $n_e$ and the strength of Ising SOC, $\lambda$. The isospin order is depicted with different colors, denoting VI (valley imbalanced), IVC (intervalley coherent), C (spin canted), and SVL (spin-valley locked) and the Fermi surface topology (`a' for annular, `s' for simply connected and `d' for disjoint) is indicated by the shade. Weakly interacting SVL regions (where the Ising spin splitting is dominated by band structure effects) are rendered in white. 
    Spin-polarized phases are present at $\lambda = 0$ but in this notation are considered to be canted phases with $\varphi =$~90$^{\circ}$.
    (\textbf{b}) Canting angle $\varphi$ of the ground state orders shown in panel \textbf{a}. Same data as main text Fig.~\ref{fig:4}a.
    }
    \label{fig:HF_Lambda_Theory}
\end{figure}

\clearpage
\section{Theory}

\subsection{Hamiltonian and methods}

We support the experimental analysis presented in this work with self-consistent Hartree-Fock simulations following the numerical implementation detailed in Ref.~\cite{koh_correlated_2023}. Our starting point is the tight-binding Hamiltonian describing bare rhombohedral trilayer graphene, which near the two valleys labeled $\tau = \pm 1$ reads
\begin{equation}
    \hat{H}_{0} = \sum_{\vb{k}} \sum_{\tau s \sigma \sigma'} 
        h(\tau \vb{K} + \vb{k})_{\sigma \sigma'} 
        c_{\tau s \sigma \vb{k}}^\dag 
        c_{\tau s \sigma' \vb{k}},
\end{equation}
where $c_{\tau s \sigma \vb{k}}$ annihilates an electron with momentum $\vb{k}$ in valley $\tau$, with spin $s \in \smash{\{\uparrow, \downarrow\}}$ and sublattice $\sigma$. The matrix $h$ contains the leading tunneling matrix elements as well as on-site potentials and reads~\cite{zhang_band_2010, zhou_half-_2021}
\begin{equation}\begin{split}
    h(\vb{q})_{\sigma \sigma'} = \mqty[
        \Delta_1 + \Delta_2 + \delta & \gamma_2 / 2 & -\gamma_0 f_{\vb{q}} & -\gamma_4 f_{\vb{q}} & -\gamma_3 f_{\vb{q}}^* & 0 \\
        \gamma_2 / 2 & \Delta_2 - \Delta_1 + \delta & 0 & -\gamma_3 f_{\vb{q}} & -\gamma_4 f_{\vb{q}}^* & -\gamma_0 f_{\vb{q}}^* \\
        -\gamma_0 f_{\vb{q}}^* & 0 & \Delta_1 + \Delta_2 & \gamma_1 & -\gamma_4 f_{\vb{q}} & 0 \\
        -\gamma_4 f_{\vb{q}}^* & -\gamma_3 f_{\vb{q}}^* & \gamma_1 & -2 \Delta_2 & -\gamma_0 f_{\vb{q}} & -\gamma_4 f_{\vb{q}} \\
        -\gamma_3 f_{\vb{q}} & -\gamma_4 f_{\vb{q}} & -\gamma_4 f_{\vb{q}}^* & -\gamma_0 f_{\vb{q}}^* & -2 \Delta_2 & \gamma_1 \\
        0 & -\gamma_0 f_{\vb{q}} & 0 & -\gamma_4 f_{\vb{q}}^* & \gamma_1 & \Delta_2 - \Delta_1]_{\sigma \sigma'}.
\end{split}\end{equation}
Here the sublattice basis is given by $(A_1, B_3, B_1, A_2, B_2, A_3)$ and $f_{\vb{q}} = \smash{e^{i q_y a / \sqrt{3}} + 2 e^{-i q_y a / 2 \sqrt{3}} \cos{\left(q_x a / 2\right)}}$. The Brillouin zone corners are labeled by $\tau \vb{K} = \tau (4 \pi / 3 a, 0)$ with $a = \SI{2.46}{\angstrom}$ the lattice constant in graphene, and $\Delta_1$ represents an interlayer potential difference generated by the applied displacement field $D$. All other parameters are taken following Ref.~\onlinecite{zhou_half-_2021}, fixed by comparing to {\it ab initio} calculations~\cite{zhang_band_2010}.

We first consider the long-range component of Coulomb interactions,
\begin{equation}
    \hat{H}_{\mathrm{C}} = \frac{1}{2A} \sum_{\vb{q}} V_{\mathrm{C}}(\vb{q}) 
        \normord{\rho(\vb{q}) \rho(-\vb{q})},
\end{equation}
where $A$ is the sample area, $\normord{}$ denotes normal ordering, $\smash{\rho(\vb{q})} = \smash{\sum_{\vb{k} \alpha} c_{\alpha \vb{k}}^\dag c_{\alpha (\vb{k} + \vb{q})}}$ is the long-wavelength part of the electronic density, and we introduced the combined flavor index $\alpha = (\tau, s, \sigma)$. We consider a Coulomb potential screened by metallic gates at a distance $d$ on both sides of the devices, with 
\begin{equation}
    V_{\mathrm{C}}(\vb{q}) = \frac{q_{\mathrm{e}}^2 }{2 \epsilon_r \epsilon_0 q} \tanh{(q d)};
\end{equation}
$\epsilon_0$ is the permittivity of free space and the dimensionless parameter $\epsilon_r$ is a relative permittivity. We phenomenologically incorporate electronic screening (beyond that provided by the graphite gates and the \hBN spacer layers)  by adjusting $\epsilon_r$ to get a reasonable fit with experimentally determined phase boundaries between various Stoner phases. We fix $\epsilon_r = 30$ in the main text figures, corresponding to slightly stronger screening compared to the value $\epsilon_r = 20$ previously used in Ref.~\cite{koh_correlated_2023}.

We separately consider the short-range components of Coulomb repulsion which enact intervalley exchange (also known as the ferromagnetic Hund's coupling), parameterized by $J_H > 0$, 
\begin{equation}
\hat{H}_{\text{V}} = \frac{J_H}{2A} \sum_{\vb{k} \vb{k}' \vb{q}} \sum_{\tau ss' \sigma \sigma'}
        \eta(\vb{q})_{\tau \sigma \sigma'}
        \normord{c_{(-\tau) s \sigma \vb{k}}^\dag c_{\tau s \sigma (\vb{k} + \vb{q})}
            c_{\tau s' \sigma' \vb{k}'}^\dag c_{(-\tau) s' \sigma' (\vb{k}' - \vb{q})} }.
            \label{eq:Hunds_coupling}
\end{equation}
Here $\eta(\vb{q})$ is a matrix of phase factors that enforces correct transformation properties under C$_3$ rotations (see Ref.~\cite{koh_correlated_2023} for details). Equation \eqref{eq:Hunds_coupling} can be approximately rewritten in terms of a coupling between spin densities in the two valleys~\cite{chatterjee_inter-valley_2022}; for $J_H>0$ it energetically favors aligning spins between the two valleys. The strength of this term is notoriously difficult to compute precisely from {\it ab initio} calculations and, furthermore, incorporates contributions from other lattice-scale interactions including the electron-phonon coupling. In this work we thus fix its value to $J_H = 4$ eV $\times A_{\rm uc} \approx 210$~meV~nm$^2$ using a phenomenological procedure that consists of reproducing the phase boundaries observed experimentally~\cite{zhou_half-_2021, arp_intervalley_2024}, as described in more detail in Ref.~\cite{koh_correlated_2023}. We note that this value is on the same order of magnitude as most theoretical estimates~\cite{alicea_graphene_2006, wei_landau-level_2024}.

Finally, we incorporate proximity-induced Ising spin-orbit coupling (SOC) described by 
\begin{equation}
    \hat{H}_{\text{I}} = \frac{\lambda}{2} \sum_{\vb{k}} 
        \vb{c}^\dag_{\vb{k}} \left( \tau^z s^z \mathbb{P}_3 \right) \vb{c}_{\vb{k}}.
\end{equation}
Here $\vb{c}^T_{\vb{k}}$ combines the relevant fermion operators, $\lambda$ denotes the Ising energy scale, $\mathbb{P}_3$ projects onto the layer of rhombohedral trilayer graphene proximal to \WSE, and the Pauli matrices $\tau^z$ and $s^z$ act on the valley and spin degree of freedom, respectively.

We implement a symmetry-restricted, self-consistent Hartree-Fock algorithm to explore the phase diagram of the system as a function of electronic density $n_e$ and interlayer potential $\Delta_1$. The details of the numerical implementation can be found in Ref.~\onlinecite{koh_correlated_2023}.

\subsection{Free energy analysis}

To understand the experimentally observed trends in a more intuitive way, we supplement the self-consistent numerical calculations with a symmetry-informed Ginzburg-Landau analysis. Following Ref.~\cite{dong_superconductivity_2024} we consider a minimal free energy density that describes the onset of Stoner-like ferromagnetism in spin-orbit-proximitized rhombohedral trilayer devices:
\begin{equation}
    {\cal F} = \frac{\kappa}{2} \sum_{\tau} \vb{n}_\tau^2 
    - J_H \vb{n}_+ \cdot \vb{n}_- + \frac{\lambda}{2} \sum_\tau \tau n^z_\tau + \cdots,
    \label{eq:free_energy_Stoner}
\end{equation}
written in terms of variables $\vb{n}_\tau$ that describe the magnitude ($|\vb{n}_\tau|$) and orientation ($\hat{\vb{n}}_\tau$) of the spin polarization in valley $\tau = \pm 1$. The first term in Eq.~\eqref{eq:free_energy_Stoner} describes the tendency towards generating a spin-polarization in each valley separately, while the second term describes the Hund's coupling between spins in the two valleys. We assume $J_H>0$ which favors the emergence of ferromagnetism. Finally, the last term describes proximity-induced spin-orbit coupling. The ellipsis denotes higher-order terms that, among other consequences, determine whether the spin-canting transition is first- or second-order.

Motivated by experiment, we now specialize to valley-balanced solutions, where the magnitudes of the spin polarization in the two valleys are equal, and define $n_p = |\vb{n}_+| + |\vb{n}_-|$. (Note that this quantity in general differs from the electronic doping level $n_e$. For a fully-polarized ferromagnetic half metal, we have $n_p = |n_e|$; in the presence of minority pockets, we instead have $n_p < |n_e|$.) The free energy in Eq.~\eqref{eq:free_energy_Stoner} then admits two different solutions that are understood intuitively as follows. For sufficiently small polarization density, the requirements of Ising SOC win over those of the Hund's coupling, due to its linear dependence on $n_p$. Indeed, for $\lambda > 2 J_{\rm H} n_p$
one finds a purely spin-valley locked phase with no net spin magnetization. In contrast, for $\lambda < 2 J_{\rm H} n_p$ one finds a spin-canted solution, depicted  schematically in Fig.~\ref{fig:3}a in the main text, whereby the spins in each valley tilt from their respective Ising axes towards the graphene plane by an angle
\begin{equation}
  \varphi = \arccos \left( {\frac{\lambda }{2 J_{\rm H} n_p}} \right).
  \label{canting_angle}
\end{equation}

This solution exhibits a non-zero magnetic moment in a spontaneously-chosen in-plane direction---thus breaking the U(1) spin-rotation symmetry about the $z$ axis. 
The corresponding in-plane magnetization $m_{\parallel}$ serves as an order parameter for the canting transition, and is given by (assuming a $g$ factor of $2$) 

\begin{equation}
    m_{\parallel} = \mu_B n_p \sin (\varphi) = \mu_B n_p \sqrt{1 - \left( \frac{\lambda}{2 J_{\rm H} n_p} \right)^2 },
\end{equation}

as in Eq.~\eqref{eq:canting_evolution} of the main text. In our Hartree-Fock simulations we numerically extract the canting angle using 

\begin{equation}
\label{eq:canting_angle}
    \varphi = \arctan \abs{ \frac{ \langle \tau^0 s^x \rangle}{ \langle \tau^z s^z \rangle}},
\end{equation}

where $\langle \hdots \rangle$ denotes an expectation value taken in the Hartree-Fock ground state.

The onset of spin canting in the density and displacement-field tuned phase diagram, as observed in Hartree-Fock simulations in Fig.~\ref{fig:4}b and c, can be understood from the free-energy perspective. The half-metal region of the phase diagram occupies a roughly diagonal region in the $\Delta_1 - n_e$ plane that tracks the behavior of the van Hove singularity. As one moves along this diagonal from $\Delta_1 = 0$, both the total density and the effective strength of interactions (through the Fermi-level density of states) increase, which has the effect of enhancing $n_p$. When $n_p$ reaches the scale $\sim \lambda/J_H$, spin canting order sets in. The ratio of Ising SOC to Hund's coupling thus set the polarization density at which canting order first appears. Upon moving further along this diagonal, the canting angle increases with $n_p$ as evident from Fig.~\ref{fig:4}c.

The same argument also explains why the quarter-metal SVL-IVC and the half-metal SVL phases respond differently to introducing Ising SOC, as shown in Fig.~\ref{fig:4}a and  Fig.~\ref{fig:HF_Lambda_Theory}. The half-metal phase generically has a higher polarization density than the quarter-metal phase, and thus it is harder to fully lock the spins back along their Ising axis. At the meV-scale values of $\lambda$ in typical spin-orbit-proximitized devices, we expect spin canting in the half-metal regions to be energetically relevant, while quarter-metal states should remain spin-valley-locked, consistent with the experimental data.

\end{document}